\documentclass[12pt]{iopart}

\usepackage{graphics}
\usepackage{graphicx}
\usepackage{dcolumn}
\usepackage{bm}
\usepackage{amsfonts}
\usepackage{bbm} 
\usepackage{iopams}
\usepackage{amssymb}
\usepackage{subfigure}

\def\ket#1{\left|#1\right>}
\def\bra#1{\left<#1\right|}

\newcommand{\bracket}[2]{\langle \, #1 \mid #2 \, \rangle}
\def\ket#1{\left|#1\right>}
\def\bra#1{\left<#1\right|}
\newcommand{\ketbra}[2]{\ket{#1}\hspace{-4pt}\bra{#2}}
\newcommand{\proj}[1]{\ketbra{#1}{#1}}

\newcommand{\abs}[1]{\mbox{$\mid #1 \mid$}}

\newcommand{\eea}{\end{eqnarray}}
\newcommand{\bea}{\begin{eqnarray}}
\newcommand{\eeas}{\end{eqnarray*}}
\newcommand{\beas}{\begin{eqnarray*}}

\newcommand{\RR}{\ensuremath{\mathbbm{R}}}

\renewcommand{\mr}[1]{\mathrm{#1}}
\renewcommand{\textrm}[1]{\mathrm{#1}}
\renewcommand{\Re}{\mathrm{Re}}
\renewcommand{\Im}{\mathrm{Im}}
\newcommand{\acosh}{\mathrm{arccosh}}

\newcommand{\text}[1]{\mathrm{#1}}
\newcommand{\openone}{\mathbbm{1}}

\begin{document}

\title{Interfacing nuclear spins in quantum dots to cavity or traveling-wave fields}

\author{Heike Schwager, J. Ignacio Cirac, and G\'eza Giedke}

\address{Max-Planck-Institut f\"{u}r Quantenoptik, D-85748
Garching,
  Germany}

\date{\today}

\begin{abstract}
We show how to realize a quantum interface between optical fields
and the
  polarized nuclear spins in a singly charged quantum dot, which is strongly
  coupled to a high-finesse optical cavity. An effective direct coupling
  between cavity and nuclear spins is obtained by adiabatically eliminating
  the (far detuned) excitonic and electronic states.\\ The requirements needed
  to map qubit and continuous variable states of cavity or traveling-wave
  fields to the collective nuclear spin are investigated: For cavity fields,
  we consider adiabatic passage processes to transfer the states. It is seen that
  a significant improvement in cavity lifetimes beyond present-day technology
  would be required for a quantum interface. We then turn to a scheme which couples
  the nuclei to the output field of the cavity and can tolerate significantly shorter cavity lifetimes.
  We show that the lifetimes reported in the literature and the recently achieved nuclear
polarization of $\sim 90$\% allow both high-fidelity read-out and
write-in of quantum information between the nuclear spins and the output field. \\
  We discuss the performance of the scheme and provide a convenient
  description of the dipolar dynamics of the nuclei for highly
  polarized spins, demonstrating that this process does not affect the
  performance of our protocol.
\end{abstract}


\maketitle
\section{Introduction}

An important milestone on the path to quantum computation and
quantum communication networks is the coupling of ``stationary''
qubits for storage and data processing (usually assumed to be
realized by material systems such as atoms or electrons) and mobile
``flying'' qubits for communication (typically photons)
\cite{Div00,Zoller2005}. Detection and subsequent storage of
information is inapplicable in quantum information as an unknown
quantum state cannot be determined faithfully by a measurement.
Hence the development of ``light-matter interfaces'' that allow the
coherent write-in and read-out of quantum information has been the
subject of intense theoretical research \cite{CZKM97,KMP99,FlLu02}.
Two paths have been identified to make light efficiently couple to a
single atomic quantum system: the use of a high-finesse cavity
coupled to a single atom or the use of an optically thick ensemble
of atoms, in whose \emph{collective} state the quantum information
is to be stored. Both have resulted in the experimental
demonstration of such interfaces
\cite{MHN+Haroche97,JSC+04,WWKR07,RBV+07,CDLK08}. Even without
strong coupling a quantum interface can be realized by combining the
probabilistic creation of entanglement between atom and light with
teleportation. This approach has been demonstrated with trapped ions
\cite{BLDM04}.

For qubits realized by electron spins in quantum dots
\cite{LoDi98,IAB+99} such interfaces have yet to be realized, though
in particular for self-assembled quantum dots \cite{HaAw08}, which
have many atom-like properties, several proposals exist to map
photonic states to an electron in a quantum dot \cite{IAB+99,YLS05}
in analogy to the atomic schemes. Strong coherent coupling between a
single quantum system and a single mode of high-Q micro- and
nano-cavities has been demonstrated experimentally
\cite{RSL+04,YSH+04,HBW+07,Reith08}, raising the prospect of
coupling light to the quantum dot's electronic state by adapting
protocols such as \cite{CZKM97}.

Despite their good isolation from many environmental degrees of
freedom, the electron-spin coherence time in today's quantum dots is
limited mainly due to strong hyperfine coupling to lattice nuclear
spins. Moreover, the capacity of such an interface is one qubit
only, making the interfacing difficult for many-photon states of the
light field as used in continuous variable quantum information
processing. In contrast, the ensemble of lattice nuclear spins could
provide a high-dimensional and long-lived quantum memory
\cite{TIL03}.

We show in the following how to couple an optical field directly to
the nuclear spin ensemble, thus interfacing light to an
exceptionally long-lived mesoscopic system that enables the storage
and retrieval of higher-dimensional states and is amenable to
coherent manipulation via the electron spin \cite{TGC+04}. The
system we consider is a charged quantum dot strongly coupled to a
high-finesse optical cavity by a detuned Raman process, introduced
in Section \ref{section1}. In Section \ref{sec:coupling} we show
that by adiabatically eliminating the trion and the electron spin
different effective couplings (that can be tuned on- and
off-resonant) between light and nuclear spins are achieved. In
Sections \ref{sec:zener} and \ref{sec:stirap}, we demonstrate that
the state of the cavity field can be directly mapped to the nuclear
spins using the methods of Landau-Zener transitions
\cite{Lan32,Ze1932} and stimulated Raman adiabatic passage (STIRAP)
\cite{BTS98}, respectively, where the latter yields a reduction of
the time required for write-in. However, the drawback of this
approach is that it requires very long cavity lifetimes. To address
this problem we discuss in Section \ref{sec:badcavity} at length an
approach that was proposed in \cite{SCG08} (here, we discuss it for
an experimentally more promising setup), which is robust against
cavity decay: the read-out maps the nuclear state to the output mode
of the cavity, while the write-in proceeds by deterministic creation
of entanglement between the nuclear spins and the cavity output-mode
and subsequent teleportation \cite{BrKi98}. In Section
\ref{sec:badcavity}, we give further insight into this system and
its dynamics: we describe the full time evolution of the system,
compute the read-out fidelity and derive the shape of the output
mode-function. Moreover, we show that apart from mapping light
states to the nuclear spins, the interaction we describe can be used
to generate an arbitrary Gaussian state. In Section \ref{sec:approx}
we discuss different aspects concerning the experimental realization
and the approximations used in our scheme such as the internal
nuclear dynamics, dominated by dipolar interactions, which we model
numerically and corrections to the first order bosonic description.

\section{The system}\label{section1}
\begin{figure}[b]
  \begin{center}
  \subfigure[][]{
    \label{systemhv1}
    \includegraphics[width=0.4\textwidth]
{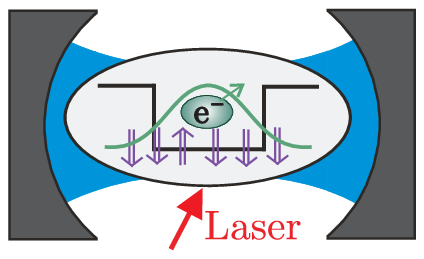}}\hfill
  \subfigure[][]{
    \label{systemhv2}
    \includegraphics[width=0.4\textwidth]{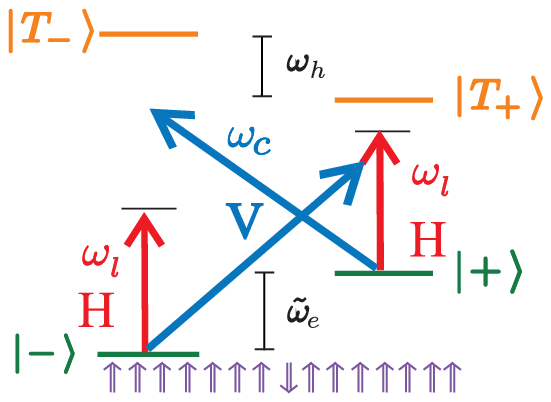}
  }
\caption{(a) A singly charged QD coupled to a high-Q
    optical cavity. (b) Level scheme of the QD. Optical and hyperfine
    driven transitions.}
 \label{fig:1}
 \end{center}
\end{figure}

We consider a self-assembled III-V quantum dot (QD) with a single
conduction-band electron strongly coupled to a high-Q nano-cavity
[see figure \ref{fig:1}a)]. At zero magnetic field, the two
electronic ground states $\ket{\pm 1/2}$ (s-type conduction band
states) are degenerate and the only dipole allowed transitions are
to the trion states $\ket{\pm 3/2}$ with spin $+3/2$ and spin $-3/2$
(heavy-hole valence band state) with $\sigma^{\pm}$ polarized light.
An external magnetic field  $B_z$ in $z$-direction, perpendicular to
the growth ($y$-) direction, Zeeman splits the two electronic states
and the trion states and leads to eigenstates
$\ket{\pm}=\frac{1}{\sqrt{2}}(\ket{1/2}\pm i\ket{-1/2})$ and
$\ket{T_{\pm}}=\frac{1}{\sqrt{2}}(\ket{3/2}\pm i\ket{-3/2})$. The
states $\ket{+}\Leftrightarrow\ket{T_+}$ and
$\ket{-}\Leftrightarrow\ket{T_-}$ can be coupled [see figure
\ref{fig:1}b)] by horizontally polarized light, and
$\ket{+}\Leftrightarrow\ket{T_-}$ and
$\ket{-}\Leftrightarrow\ket{T_+}$ by vertically polarized light:
\begin{eqnarray}\label{eqn:hamiltonoptischanfang}
H_{\text{opt}}=&\frac{\Omega_c}{2}\,(a^{\dagger}
\ketbra{-}{T_+}+a^{\dagger}\ketbra{+}{T_-}+\textrm{h.c.})\nonumber\\&+\frac{\Omega_l}{2}\,(e^{+i\omega_l
t}(\ketbra{+}{T_+}+\ketbra{-}{T_-})+\textrm{h.c.})\nonumber\\
&+\omega_c\,a^{\dagger}a+\omega_{T_+}\ketbra{T_+}{T_+}+\omega_{T_-}\proj{T_-}
+\omega_{e}S^z.
\end{eqnarray}
Here, $\hbar=1$, $S^z$ is the electron spin operator, $a^{\dagger}$
and $a$ are creation and annihilation operators of the single mode
cavity field and $\omega_c$, $\omega_l$ denote the cavity and the
laser frequency (which are vertically/horizontally polarized
respectively) and $\Omega_c$, $\Omega_l$ the Rabi frequencies of the
cavity and the laser field, respectively. The energies of the trion
states $\ket{T_+}$, $\ket{T_-}$ are
$\omega_{T_+}=\omega_T+\omega_{h}/2$ and
$\omega_{T_-}=\omega_T-\omega_{h}/2$ where $\omega_T$ is the energy
of the trion (without magnetic field), $\omega_{h}$ the energy of
the hole Zeeman splitting and $\omega_{e}=g_e\,\mu_b B_{y}$ denotes
the Zeeman splitting of the electronic states. The first term of the
Hamiltonian given by (\ref{eqn:hamiltonoptischanfang}) describes the
coupling to the cavity field and the second term the coupling to a
classical laser field. We assume both cavity decay and spontaneous
emission rate of the QD to be much smaller than $\Omega_c$ and omit
both processes in (\ref{eqn:hamiltonoptischanfang}). Besides the
coupling to optical fields, the electron spin in a QD also has a
strong hyperfine interaction with the lattice nuclear spins, which
is for s-type electrons dominated by the Fermi contact term
\begin{equation}\label{eqn:hfneu}
H_{\text{hf}}=\frac{A}{2}(A^+S^-+S^+A^-) +  AS^zA^z,
\end{equation}
where $S^{\pm,z}$ are the electron spin operators and
$A^{\pm,z}=\sum_j\,\alpha_j I_j^{\pm,z}$ are the collective nuclear
spin operators (in a typical GaAs quantum dot, the number of
Ga and As nuclei lies between $N\sim10^4$-$10^6$). The individual
coupling constants $\alpha_j$ are proportional to the electron wave
function at site $j$ (and the magnetic dipole moment of the $j$th
nucleus) \cite{SKL03} and are normalized to $\sum_j\alpha_j=1$.
The requirement for using nuclear spins as a quantum memory is to
initialize them in a well-defined, highly polarized state. By this
we mean that $\left<A^z\right>$ is close to its minimum value
$\left<A^z\right>_{\textrm{min}}$ ($\approx-1/2$ for spin-1/2
nuclei) and define the polarization as
$P=\left<A^z\right>/\left<A^z\right>_{\textrm{min}}$. Due to their
small magnetic moments, nuclear spins are almost fully mixed even at
dilution-fridge temperatures and fields of several Tesla. Over the
past years, large progress in dynamical polarization experiments
 \cite{BSG+05,MBI07,SNM+08,UBM+07} has been reported with nuclear polarization up to $60\%$,
 recently, nuclear polarization
$>80$\% has been achieved \cite{Maletinsky2008}.

A convenient and intuitive description of the highly polarized
nuclei with homogeneous coupling to the electron is provided by the
Holstein Primakoff transformation \cite{HoPr40}, by which collective
nuclear spin operators $A^{\pm,z}$ can be mapped to the bosonic
operators $b$,$b^{\dagger}$, associating
$A^-\rightarrow\frac{1}{\sqrt{N}}\sqrt{1-\frac{b^{\dagger}b}{N}}\,b$
and
$A^z\rightarrow\frac{1}{N}\left(b^{\dagger}b-\frac{N}{2}\right)$.
Assuming high polarization, the electron spin couples to a bosonic
``spin wave'' described by $ A^-= \frac{1}{\sqrt{N}}b$ and
$A^z=\frac{1}{N}(b^\dagger b-N/2)$ by a Jaynes-Cummings-like
interaction
\begin{eqnarray}\label{eqn:hfbosonic}
H_{\text{hf}}&=&\frac{g_n}{2}(b^{\dagger}S^-+S^+b) +
\frac{g_n}{\sqrt{N}}S^z\left(b^{\dagger}b-\frac{N}{2}\right) ,
\end{eqnarray}
with $g_n=A/\sqrt{N}$. The initial state of the nuclear spins is
represented by a collection of bosonic modes, with $b$ in the vacuum
state. One can generalize this description to the case of
inhomogeneous coupling to the electron ($g_n=A\sqrt{\sum_i
\alpha_i^2}$) and obtains an identical description in $0$th order in
$\left<\frac{b^{\dagger}b}{N}\right>=(1-P)/2$ \cite{CCG07b}.
Corrections to this description arising from inhomogeneous coupling
and not fully polarized nuclear spins will be discussed briefly in
Section \ref{sec:errorbos} and a detailed discussion can be found in
\cite{Christ2008}. It should be noted that the scheme we present
does not \emph{require} the bosonic description and could also be
discussed directly in terms of the collective spin operators. The
Fock basis would be replaced by
$(A^+)^n\ket{\downarrow\dots\downarrow}$ and errors due to the
inhomogeneity would have to be treated along the lines of
\cite{TIL03} and \cite{SCG08}. The bosonic picture, however, allows
a much more transparent treatment of the corrections to the ideal
case, emphasizes the relation to quantum optical schemes, and gives
access to the Gaussian toolbox of entanglement criteria and
transformations.

%

\section{Effective coupling between nuclei and
cavity}\label{sec:coupling}

Our aim is to obtain from $H=H_\mr{opt}+H_\mr{hf}$ a direct coupling
between nuclear spins and light. The Hamiltonian $H$ describes a
complicated coupled dynamics of cavity, nuclei and quantum dot.
Instead of making use of the full Hamiltonian (and deriving the
desired mapping, e.g., in the framework of optimal control theory)
we aim for a simpler, more transparent approach. Closely following
\cite{SCG08}, we adiabatically eliminate \cite{BPM06} the trion and
the electronic spin degree of freedom, which leads to a Hamiltonian
$H_{\mathrm{el}}$ that describes a direct coupling between nuclear
spins and light. As explained later, this can be achieved if the
couplings (the Rabi frequency of the laser/cavity, the hyperfine
coupling, respectively) are much weaker than the detunings to the
corresponding transition:
\numparts
  \begin{eqnarray}
&\Delta'_{T_{\pm}}\gg\Omega_l,\Omega_c\sqrt{n},\label{eq:condelimi}\\
&\sqrt{\Delta'_{T_{\pm}}\,\,\tilde{\omega}_e}\gg\Omega_l,\Omega_c\sqrt{n},\label{eq:condelimii}\\
&\tilde{\omega}_e\gg g_n\sqrt{m}.\label{eq:condelimiii}
  \end{eqnarray}
\endnumparts
Here, $\Delta'_{T_{\pm}}=\omega_{T}-\omega_l\pm\omega_{h}/2
+\tilde{\omega}_{e}/2=\Delta'\pm\omega_{h}/2+\tilde{\omega}_e/2$
with $\Delta'=\omega_{T}-\omega_l$, $n$ is the number of cavity
photons and $m$ the number of nuclear excitations. Note that
typically $\tilde\omega_e<\Delta'_{T_{\pm}}$ such that condition
(\ref{eq:condelimi}) becomes redundant. In addition to
(\ref{eq:condelimi})-(\ref{eq:condelimiii}), we choose the
adjustable parameters such that all first order and second order
processes described by $H$ are off-resonant, but the (third order)
process in which a photon is scattered from the laser into the
cavity while a nuclear spin is flipped down (and its converse) is
resonant. This leads to the desired effective interaction.

The idea of adiabatic elimination is to perturbatively approximate a
given Hamiltonian by removing a subspace from the description that
is populated only with a very low probability due to chosen initial
conditions and detunings or fast decay processes. If initially
unpopulated states (in our case the trion state $\ket{X}$ and the
electronic spin-up state $\ket{\uparrow}$) are only weakly coupled
to the initially occupied states, they remain essentially
unpopulated during the time evolution of the system and can be
eliminated from the description. The higher order transitions via
the eliminated levels appear as additional energy shifts and
couplings in the effective Hamiltonian on the lower-dimensional
subspace.

The starting point is the Hamiltonian
$H=H_{\mathrm{opt}}+H_{\mathrm{hf}}$ given by
(\ref{eqn:hamiltonoptischanfang}) and (\ref{eqn:hfneu}). In order to
get a time-independent Hamiltonian, we go to a frame rotating with
$U^{\dagger}=\exp{[-i\omega_lt(a^{\dagger}a+
\proj{T_+}+\proj{T_-})]}$ and obtain:
\begin{eqnarray}\label{rotatingframeham}
H'&=&\frac{\Omega_c}{2}\,(a^{\dagger}
\ketbra{-}{T_+}+a^{\dagger}\ketbra{+}{T_-}+\textrm{h.c.})
 +\frac{\Omega_l}{2}\,(\ketbra{+}{T_+}+\ketbra{-}{T_-})+\textrm{h.c.})\nonumber\\&&+\delta\,
a^{\dagger}a+\Delta_{T_+}\proj{T_+}+\Delta_{T_-}\ketbra{T_-}{T_-}+\omega_{e}S^z+H_{hf},
\end{eqnarray}
where $\Delta_{T_\pm}=\omega_{T_\pm}-\omega_l$ and
$\delta=\omega_c-\omega_l$.

Choosing the cavity and laser frequencies, $\omega_c$ and
$\omega_l$, far detuned from the exciton transition and the
splitting of the electronic states $\tilde{\omega}_e$ much larger
than the hyperfine coupling $g_n$, such that conditions
(\ref{eq:condelimi})-(\ref{eq:condelimiii}) are fulfilled, we can
adiabatically eliminate the states $\ket{T_{\pm}}$ and $\ket{+}$. A
detailed derivation of the adiabatic elimination can be found in
\cite{SCG08}. It yields a Hamiltonian, that describes an effective
coupling between light and nuclear spins
\begin{eqnarray}
  \label{eq:Heff1a}
  H_{\mathrm{el}} =&\frac{\Omega_c\Omega_lA}{8\Delta'_{T_+}\tilde{\omega}_e}(aA^+
  +\mr{h.c.})+\frac{\Omega_c\Omega_lA}{8\Delta'_{T_-}\tilde{\omega}_e}(aA^-
  +\mr{h.c.})+\omega_1 a^{\dagger}a-\frac{A}{2}  \delta
A^z\nonumber\\&-\frac{A^2}{4\tilde\omega_e}A^+A^-+T_{\mathrm{nl}},
\end{eqnarray}
where the energy of the photons
$\omega_1=\delta-\frac{\Omega_c^2}{4{\Delta'_{T_+}}}+\frac{\Omega_l^2}{4\Delta'^2_{T_-}}\delta$
and the energy of the nuclear spin excitations $\sim
-\frac{A}{2N}-\frac{A^2}{4N\tilde{\omega}_e}$. By $T_{\mathrm{nl}}$
we denote the nonlinear terms
\begin{equation}
T_{\mathrm{nl}}=\frac{A^3}{8\tilde{\omega}_e^2} A^+\delta
A^zA^-+\frac{A^2}{4\tilde\omega_{e}^2}\delta
a^{\dagger}aA^+A^-+\frac{\Omega_c^2\delta}{4{\Delta'^2_{T_+}}}a^{\dagger}a^{\dagger}aa,
\end{equation}
which are small
($\|T_{\mathrm{nl}}\|\ll\frac{\Omega_c\Omega_lA}{8\Delta'\tilde{\omega}_e}
$) in the situation we consider ($\delta\ll\Omega_c,
g_n/\tilde{\omega_z}\sim\Omega_l/\Delta'_{T_+,T_-}\ll1$) and
neglected in the following. We also neglect the nuclear Zeeman term
which is of order $10^{-3}$ smaller than the Zeeman energy of the
electron. In the bosonic description of the nuclear spins that we
introduced in (\ref{eqn:hfbosonic}) the Hamiltonian given by
(\ref{eq:Heff1a}) reads
\begin{eqnarray}
  \label{eq:Heff1b}
  H_{\text{eff}} = g_1 (ab^{\dagger}
  +\mr{h.c.})+g_2(ab
  +\mr{h.c.})+\omega_1 a^{\dagger}a+\omega_2
  b^{\dagger}b,
\end{eqnarray}
with coupling strengths $g_1$ and $g_2$ given by
\begin{equation}\label{eq:gideal}
g_{1}=\frac{\Omega_c\Omega_lg_n}{8\Delta'_{T_{+}}\tilde{\omega}_e},\,\,\,\,\,\,
g_{2}=\frac{\Omega_c\Omega_lg_n}{8\Delta'_{T_{-}}\tilde{\omega}_e}.
\end{equation}
The energy of the nuclear spin excitations can now be written as
$\omega_2=-\frac{A}{2N}-\frac{g_n^2}{4\tilde{\omega}_e}$. The first
term in the Hamiltonian is a beamsplitter type interaction $\sim
(ab^{\dagger}+\mr{h.c.})$ whereas the second term is a two-mode
squeezing type interaction $\sim (ab
  +\mr{h.c.})$. Both interactions can be made dominant by choosing the
  resonance condition to be either $\omega_1=\omega_2$ or $\omega_1=-\omega_2$.
This will be discussed in detail in the following and illustrated
numerically.

\begin{figure}[htbp]
  \centering
  \subfigure[][]{
    \label{squeezing_bs1.eps}
    \includegraphics[width=0.48\textwidth]
{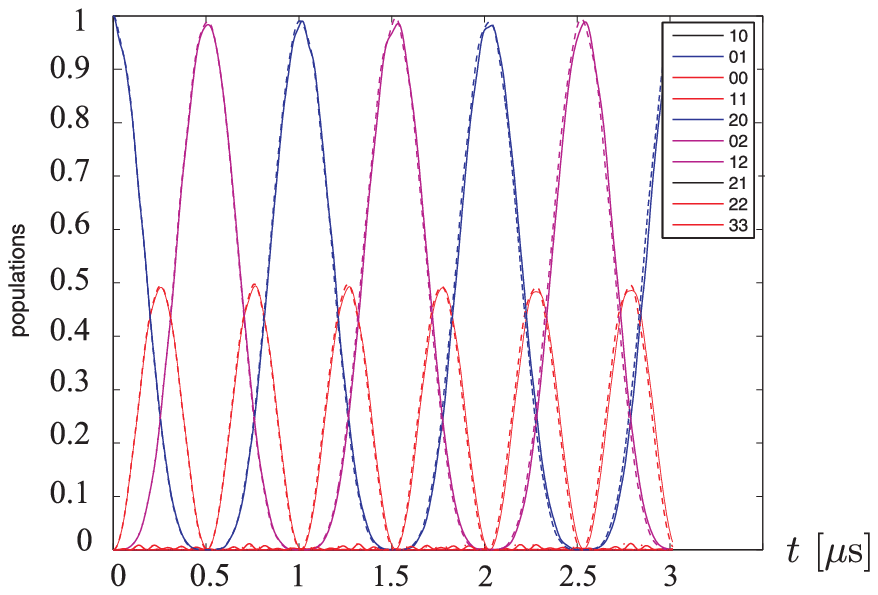}}\hfill
  \subfigure[][]{
    \label{squeezing_bs2.eps}
    \includegraphics[width=0.48\textwidth]{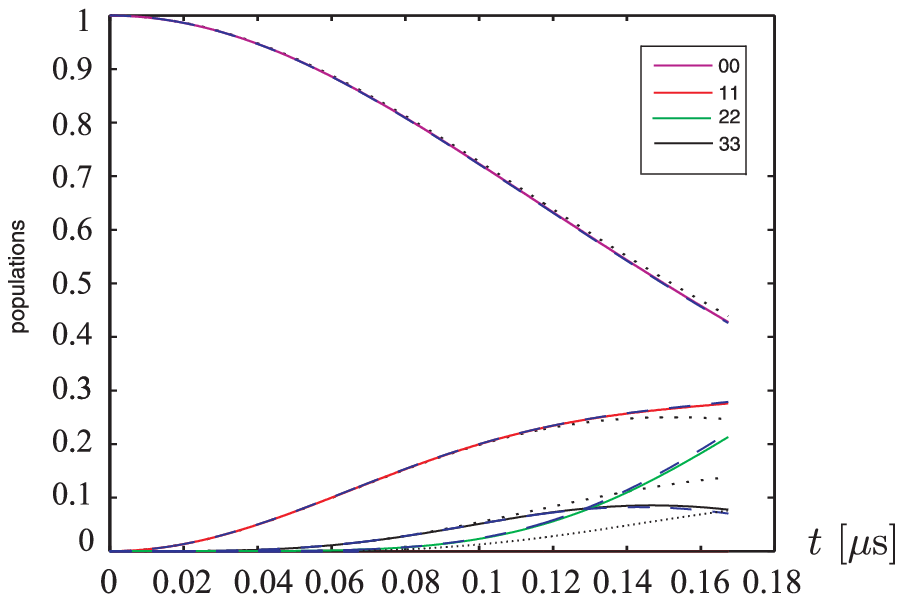}
  }
  \caption{a) Evolution of the two-photon Fock state $\psi_{20}$ under the full Hamiltonian
  $H'$ (solid lines)
  and the eliminated Hamiltonian $H_{\text{eff}}$ (dashed lines) tuning the energies such that $\omega_1=\omega_2$ (beamsplitter-type interaction).
  b) Evolution of the state $\psi_{00}$ under the full Hamiltonian
  $H'$ (solid lines)
  and the eliminated Hamiltonian $H_{\text{eff}}$ (dashed blue lines) tuning the energies such that $\omega_1=-\omega_2$ (squeezing interaction).
  The dotted black lines show the evolution under exact two-mode squeezing (up to $n=3$).}
  \label{fig:elimination}
\end{figure}

First, we validate the adiabatic elimination by a numerical
simulation which compares the evolution of states $\Psi_{20}$ (where
the first subscript indicates the number of photons and the second
the number of nuclear excitations) [under the condition
$\omega_1=\omega_2$, see figure \ref{fig:elimination}a)] and
$\Psi_{00}$ [under the condition $\omega_1=-\omega_2$, see figure
\ref{fig:elimination}b)] under the full Hamiltonian given by
(\ref{rotatingframeham}) to the evolution under the eliminated
Hamiltonian given by (\ref{eq:Heff1b}). The solid lines show the
evolution under the full Hamiltonian $H'$, the dashed lines under
the eliminated Hamiltonian $H_{\text{eff}}$ and we find that $H'$ is
well approximated by $H_{\text{eff}}$, and that the nonlinear terms
$T_{\mathrm{nl}}$ can indeed be neglected.

For the simulation, we choose the parameters as follows: we assume a
hole g-factor $g_h=-0.31$ and an electron g-factor $g_e=0.48$
\cite{XWS+07}; the number of nuclei $N=10^4$, the hyperfine coupling
constant $A=100\mu eV$, the laser and cavity Rabi frequency
$\Omega_c=\Omega_l=15\mu eV$, the detuning of the trion
$\Delta'=1000\mu eV$, the effective Zeeman splitting of the
electronic states $\tilde{\omega}_{e}=13.9\mu eV$ (the magnetic
field in $x$-direction is $4$T) and the Zeeman splitting of the hole
$\omega_{h}=-71.8 \mu eV$. With these parameters, the conditions
given by (\ref{eq:condelimi})-(\ref{eq:condelimii}) are fulfilled
and values $g_{1}=2.1\cdot10^{-3}\mu$eV and
$g_{2}=1.9\cdot10^{-3}\mu$eV are obtained. We assume full nuclear
(spin-down) polarization and use the bosonic description.

As already mentioned, two distinct resonance conditions are chosen
in figures \ref{fig:elimination} a) and b), leading to different
dynamics of the system:

For resonant exchange of excitations between the two systems, we
choose $\omega_1=\omega_2$, where the tuning can be done by changing
$\delta=\omega_c-\omega_l$. Then $H_{\text{eff}}$ describes a
beamsplitter-like coupling of the modes $a$ and $b$ and the
effective interaction is described by
\begin{eqnarray}
  \label{eq:Hbs}
  H_{\text{bs}} = g_1 (ab^{\dagger}
  +\mr{h.c.})+\omega_1 a^{\dagger}a+\omega_2
  b^{\dagger}b.
\end{eqnarray}
Processes in which absorption (or emission) of a cavity photon is
accompanied by a nuclear spin excitation are resonant, whereas the
squeezing interaction given by $g_2 (a^\dagger b^\dagger +a b)$ is
off-resonant. This can be seen going to a frame rotating with
$\omega_1$: $g_2$ is rotating with $2\omega_1$ and as $2\omega_1\gg
g_2$, the squeezing type interaction is off resonant.

Tuning the energies such that $\omega_1=-\omega_2$, the creation of
a nuclear spin excitation is accompanied by scattering of a laser
photon \emph{into} the cavity, i.e. the effective coupling becomes
$g_2(a^\dag b^\dag+ab)$ and the beamsplitter-type interaction $g_1
(ab^{\dagger}
  +a^{\dagger}b)$ is
off-resonant. The driving laser now facilitates the \emph{joint}
creation (or annihilation) of a spin excitation and a cavity photon,
realizing a two-mode squeezing Hamiltonian
\begin{equation}
  \label{eq:Hsq}
  H_{\text{sq}} = g_2 (a^\dagger b^\dagger +a b) + \omega_1 a^\dagger a  + \omega_2
  b^\dagger b.
\end{equation}
The plots in figures~\ref{fig:elimination}a) and b) illustrate that
the dynamics of the system can indeed be approximated by
(\ref{eq:Hbs}) and (\ref{eq:Hsq}). To simulate the beamsplitter type
coupling given by (\ref{eq:Hbs}), we choose $\omega_1=\omega_2$ and
let the two-photon Fock state $\psi_{20}$ evolve under the
Hamiltonian given by (\ref{eq:Heff1b}) [see dashed lines in figure
\ref{fig:elimination} a)]. Almost perfect Rabi-oscillations can be
seen between the two-photon Fock state $\psi_{20}$ and the state
with two nuclear spin excitations $\psi_{02}$, showing that
$g_2(a^{\dagger}b^{\dagger} +\mr{h.c.})$ in (\ref{eq:Heff1b}) can
indeed be neglected. To simulate the squeezing-type interaction we
choose $\omega_1=-\omega_2$ and study the evolution of the state
$\psi_{00}$ under the Hamiltonian given by (\ref{eq:Heff1b}) [see
dashed blue lines in figure \ref{fig:elimination} b)]. It can be
seen that the state $\psi_{00}$ evolves into the states $\psi_{11}$,
$\psi_{22}$ and $\psi_{33}$ with coupling strengths $g_2\sqrt{n}$,
depending on the number of excitations $n$. We have thus shown, that
in this case, the beamsplitter-type interaction can indeed be
neglected. For simplicity, we restricted the number of photons and
nuclear excitations to $3$ in our simulation, such that states
$\psi_{44}$ and higher excitation states do not occur and the
evolution of the states $\psi_{22}$ and $\psi_{33}$ does only
correspond to its evolution in a space with higher excitation
numbers at very short times. This can be seen comparing the
evolution to the exact two-mode squeezing which generates the state
$\sqrt{1-\tanh^2{(g_2 t)}}\sum_{n=0}^{\infty}\tanh^n{(g_2
t)}\ket{nn}$ for which the populations up to $n=3$ are plotted in
figure \ref{fig:elimination} b) (dotted black lines).

\section{Landau-Zener transitions}\label{sec:zener}
To map the state of the cavity to the nuclear spins, we take
advantage of a formal analogy between the linear two-mode
interaction given by (\ref{eq:Hbs}) in the Heisenberg picture and
the Landau-Zener problem \cite{Lan32, Ze1932}. In the conventional
Landau-Zener problem, initially uncoupled Hamiltonian eigenstates of
a two-level system interact at an avoided crossing. This interaction
is achieved by slowly changing an external parameter such that the
level separation is a linear function of time. If the system starts
in the ground state, the probability of finding it in the excited
state is given by the Landau-Zener formula \cite{Lan32,Ze1932}.

Here, we invoke this idea in the Heisenberg picture to achieve a
mapping of the photon annihilation operator $a$ to the collective
nuclear spin operator in the bosonic approximation $b$, i.e.,
$a\rightarrow b$ (the bosonic operators $a$ and $b$ are initially
uncoupled). In the following we show that our system can be
transformed to a system which corresponds to the standard
Landau-Zener problem.

In the Heisenberg picture, the linear two-mode interaction between
the cavity mode and the nuclear spins in the quantum dot, given by
(\ref{eq:Hbs}), is described by a set of coupled differential
equations for the mode operators:
\begin{equation}\label{eq:lz}
\frac{\textrm{d}}{\textrm{dt}} \left(\begin{array}{c}
a(t)\\
b(t)
\end{array}\right)=-i
\left(\begin{array}{cc}
\omega_1&g_1\\
g_1&\omega_2
\end{array}\right)
\left(\begin{array}{c}
a(t)\\
b(t)
\end{array}\right).
\end{equation}
The available control parameters used to effect the change of
$\omega_1-\omega_2$ are the laser Rabi frequency $\Omega_l$ and the
laser frequency $\omega_l$. We consider a linear time dependence of
\begin{equation}\label{eq:lz2}
\omega_1-\omega_2=\beta t
\end{equation}
 for simplicity while the coupling $g_1$ is
constant.

Denoting by $a({u \choose  v})$ the operator $u a_{-\infty} +
vb_{-\infty}$ (a normalized linear combination of the purely
photonic $a_{-\infty}$ and purely nuclear $b_{-\infty}$) and by
$a_t({u \choose  v})\equiv a({u_t \choose  v_t})$ its image under
time evolution, the two operator equations of (\ref{eq:lz}) can be
combined into
\[
\frac{d}{dt}a_t({u \choose  v}) = -i a_t\left[ \left(
\begin{array}{cc}
    \omega_1&g_1\\ g_1&\omega_2
  \end{array}
 \right){u\choose  v}
\right],
\]
which can be written completely in terms of the mode function:
\begin{equation}
  \label{eq:1}
\frac{d}{dt}  {u_t\choose  v_t}=-i\left(   \begin{array}{cc}
    \omega_1&g_1\\ g_1&\omega_2
  \end{array}
 \right){u_t\choose  v_t}.
\end{equation}
This is the same kind of coupled differential equation that is
encountered in the Landau-Zener problem \cite{Lan32, Ze1932}.
Following the calculation as done in \cite{Lan32, Ze1932} we find
that
\begin{equation}\label{eqn:ze}
|v(t=\infty)|^2=K^2U_1(\infty){U_1^*(\infty)}=1-\epsilon^2,
\end{equation}
where $\epsilon\equiv e^{-\pi\gamma_z}$ and
\begin{equation}
\lim_{t\to\infty}U_1(t)=-K\frac{\sqrt{2\pi}}{\Gamma(i\gamma_z+1)}e^{-\frac{1}{4}\pi
\gamma_z}e^{i\beta t^2}(\sqrt{\beta}t)^{i\gamma_z},
\end{equation}
with $\gamma_z=\frac{g_1^2}{\beta}$
 and the constant
$K=\sqrt{\gamma_z}\exp{\left(-\frac{\gamma_z\pi}{4}\right)}$ (for a
detailed discussion see \ref{Appendix}). We thus find that the
cavity mode operator $a$ is mapped to
\begin{equation}
a_{+\infty}=\sqrt{1-\epsilon^2}\,b_{-\infty}+ \epsilon\,a_{-\infty},
\end{equation}
a dominantly nuclear operator for $\beta$ small enough so that
$\epsilon$ is small, which effectively means for large enough times,
as $\omega_1-\omega_2\equiv\beta t$.\\

\subsection{Quality of the mapping for Fock and coherent
states}\label{sec:lzfock} In the following we will consider the
quality of the mapping within the model given by
(\ref{eqn:hamiltonoptischanfang}) and (\ref{eqn:hfbosonic}), other
imperfections will be discussed in Section \ref{sec:approx}. To
evaluate the quality of the mapping, we return to the
Schr\"{o}dinger picture. The mapping of a $n$ photon Fock state of
the cavity to the nuclei leads to a mixture of Fock states with
photon numbers $\leq n$. The fidelity with which a $n$-photon Fock
state is mapped, is given by
\begin{equation}\label{eqn:fidfock}
F_{\text{F}}(n)=\bra{n}\text{tr}_{\text{c}}\left(\left[\frac{a_{\infty}^{\dagger}}{\sqrt{n!}}\right]^n\proj{0}\left[\frac{a_{\infty}}{\sqrt{n!}}\right]^n\right)\ket{n
}=(1-\epsilon^2)^n,
\end{equation}

\begin{figure}[t]
\begin{center}
\includegraphics[scale=1.5]{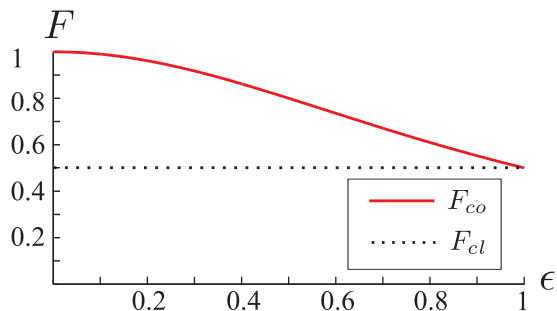}
\caption{The (solid) red line indicates the fidelity of the mapping
of an amplified coherent state of light to the nuclear spins vs the
losses of the mapping $\epsilon$. The fidelity is, even for large
losses, higher than the fidelity $F_{\text{cl}}$ that can be
achieved by classical means, indicated by the (dotted) black
line.}\label{fig:fidelitycohstate}
\end{center}
\end{figure}
where the cavity mode is traced out. In a next step, we want to know
which fidelities can be achieved for superpositions of number
states, e.g., for coherent states.

Coherent states are representatives of the family of Gaussian
states, which play an important role in quantum optics and quantum
information processing.  A Gaussian state is fully characterized by
its first and second moments $(\gamma,d)$, where $\gamma$ is the
state's covariance matrix and $d$ its displacement (see
\ref{Appendix1}). Since the dynamics generated by
(\ref{eqn:fidfock}) is Gaussian, the mapping can be fully
characterized in terms of covariance matrices. The mapping of a
Gaussian state of light onto the nuclear spins corresponds to
\begin{equation}\label{eqn:ampmap}
\left(\gamma_c,d_c\right)_{\text{c}}\stackrel{\textrm{map}}{\longrightarrow}\left((1-\epsilon^2)\gamma_c+\epsilon^2\gamma_{ns}\,
,\sqrt{1-\epsilon^2}d_{c}+\epsilon d_{ns}\right)_{\text{ns}},
\end{equation}
where $(\gamma_c, d_{c})_{\text{c}}$ and $(\gamma_{ns},
d_{ns})_{\text{ns}}$ describe the states of cavity and nuclear
spins, respectively. For a coherent state mapped to the nuclei, this
corresponds to the map
\begin{equation}
\left(\openone,\alpha\right)_{\text{c}}\stackrel{\textrm{map}}{\longrightarrow}\left(\openone,\sqrt{1-\epsilon^2}\alpha\right)_{\text{ns}}.
\end{equation}
The fidelity of the mapping is given by \cite{SZ97}
\begin{eqnarray}
F_c=|\bracket{\alpha}{\sqrt{1-\epsilon^2}\alpha}|^2=\exp\left[-\left|\left(1-\sqrt{1-\epsilon^2}\right)\alpha\right|^2\right].
\end{eqnarray}
The minimal goal of a quantum interface is to achieve a better
fidelity than can be achieved by classical means. As proved in
\cite{HWPC05, FSB+98} the classical benchmark fidelity of coherent
states distributed in phase space according to
$p(\alpha)=\frac{\lambda}{\pi}\exp{(-\lambda|\alpha|^2)}$ is given
by $F_{max}=\frac{1+\lambda}{2+\lambda}$. Averaging $F_c$ over all
possible coherent input states with a Gaussian distribution the
fidelity reads
\begin{eqnarray}
\overline{F_c}=&\int\textrm{d}^2\alpha\,p(\alpha)F_c=\frac{\lambda}{1-\epsilon^2/2-\sqrt{1-\epsilon^2}+\lambda}.
\end{eqnarray}
For a flat distribution with $\lambda\rightarrow0$ large photon
numbers that lead to high losses are dominant and therefore
$\overline{F_c}\rightarrow 0$.

A way to improve the average fidelity is to amplify the coherent
state either at the write-in or the read-out stage, thus
compensating losses due to the imperfect mapping. Optimal phase
insensitive amplification would map
$\left(\gamma,d\right)\longrightarrow\left(\kappa^2\gamma+(\kappa^2-1)\openone,\kappa
d\right)$ \cite{Cav82}. Choosing $\kappa$ such that
$\kappa\stackrel{!}{=}\frac{1}{\sqrt{1-\epsilon^2}}$, amplification
and subsequent mapping can be written as
\begin{equation}
\left(\openone,\alpha\right)_{\text{c}}\stackrel{\textrm{amp.}}{\longrightarrow}
\left(\kappa\,
\openone,\frac{1}{\sqrt{1-\epsilon^2}}\alpha\right)_{\text{c}}\stackrel{\textrm{map}}{\longrightarrow}\left(\kappa(1-\epsilon^2)+\epsilon^2)\,
\openone,\alpha\right)_{\text{ns}}.
\end{equation}
where subscript $c$ refers to the cavity and $ns$ to the nuclear
spins. The fidelity of the mapping of the amplified state $\rho_m$,
calculated using the relations for transition amplitudes of Gaussian
states in \cite{Scu98}, is given by
\begin{eqnarray}
F_{\text{co}}&=\bra{\alpha}\rho_{\text{m}}(\alpha)\ket{\alpha}=\det\left(\frac{\gamma_{c}+\gamma_{ns}}{2}\right)^{-1/2}e^{-(d_{ns}-d_c)^T(\gamma_{c}+\gamma_{ns})(d_{ns}-d_c)}\nonumber\\
&=\det\left(\frac{\kappa(2-\epsilon^2)+\epsilon^2}{2}\openone\right)^{-1/2}=\frac{1}{1+\epsilon^2}
\end{eqnarray}
\begin{figure}[b]
  \centering
  \subfigure[][]{
    \label{setuppaper1}
    \includegraphics[width=0.4\textwidth]
{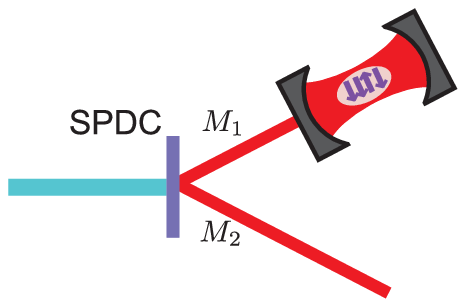}}\hfill
  \subfigure[][]{
    \label{setuppaper2}
    \includegraphics[width=0.4\textwidth]{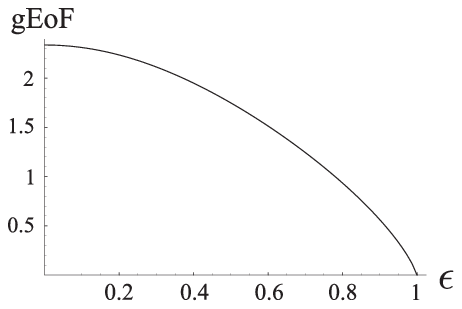}}
  \caption{a) One part ($M_1$) of a two-mode squeezed state, arising
from a Spontaneous Parametric Down Conversion source is coupled into
the cavity and mapped to the nuclear spins of the quantum dot.
Thereby, the other part ($M_2$) gets entangled with the nuclear
spins. b) Plot of the Gaussian entanglement of formation (gEoF) for
squeezing parameter $r=1$ vs mapping error $\epsilon$. The
entanglement of the nuclear spins with part of a two-mode squeezed
state ($M_2$) is a decreasing function for increasing mapping error
$\epsilon$.}
 \label{setuppaper}
\end{figure}

A plot of $F_{\text{co}}$ is shown in figure
\ref{fig:fidelitycohstate}. The fidelity of the mapping of the
amplified coherent state is always higher than the classical
benchmark fidelity $F_{\text{cl}}=\frac{1}{2}$ for a flat
distribution i.e., the quantum interface shows high
performance even for large losses $\epsilon$.\\

\subsection{Storage of an entangled state}\label{sec:lzent}

Up to now we have shown that it is possible to transfer Fock and
coherent states of light onto the nuclear spin memory. However, the
ultimate test for a quantum memory is whether it is capable of
faithfully storing part of an entangled quantum system. As an
example for an entangled light state we consider a two-mode squeezed
state where one of its light modes, $M_1$, is coupled into the
cavity and mapped onto the nuclear spins of the quantum dot [see
figure \ref{setuppaper}a)]. To see how well the entanglement is
preserved, we compute the entanglement between the nuclear spins and
the light mode $M_2$ using Gaussian entanglement of formation
\cite{WGK+03}. We find it to be a monotonically decreasing function
of the mapping error $\epsilon \in [0,1]$. As seen from figure
\ref{setuppaper}b) the nuclear spins of the quantum dot are
entangled with the light mode $M_2$. This allows a remote access to
the memory via teleportation,
required for e.g., quantum repeaters.\\

\subsection{Mapping time}

A timescale for the mapping can directly be found considering
Hamiltonian $H_\text{bs}$ where the parameters are chosen such that
$\omega_1=\omega_2$ and set to zero in a rotating frame: acting for
a time $t=\pi/g$ (which is for the parameters used in Section
\ref{section1} on the order of $6 \mu$s) it maps $a\to b$ and $b\to
a$ thus realizing a swap gate between cavity and nuclear spins. This
setting however, would in contrast to the adiabatic methods
discussed in this and the following Sections, be sensitive to timing
errors because letting $H_{\text{bs}}$ act for too long would
reverse the mapping.

\section{STIRAP}\label{sec:stirap}

Mapping the state of the cavity to the nuclear spins is also
possible considering a system where only the trion states are
adiabatically eliminated and elimination of the electronic states is
not required to achieve the desired interaction.
\begin{figure}[t]
  \centering
  \subfigure[][]{
    \label{stirap1}
    \includegraphics[width=0.3\textwidth]
{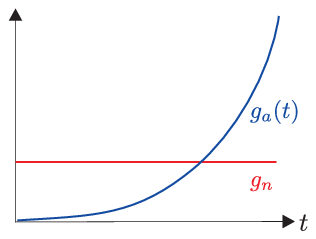}}\hfill
  \subfigure[][]{
    \label{stirap2}
    \includegraphics[width=0.49\textwidth]{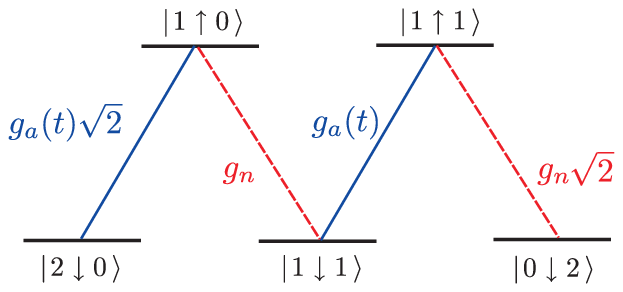}}
  \caption{a) Schematic view of the time dependence of the coupling
$g_a$ and the constant hyperfine coupling $g_n$. b) Schematic view
of the level scheme of the system with initially two photons. The
hyperfine coupling $g_n$ is "always on", whereas $g_a(t)$ is an
increasing function for which $g_a(T)\gg g_n$.}
 \label{fig:stirap3}
\end{figure}
We show that, with this system, the process of storing a state of
light to the nuclear spins can be achieved by the well-known
technique of Stimulated Raman Adiabatic Passage
(STIRAP)\cite{BTS98}, which has been studied for multilevel systems
\cite{PMZ+95} and has been demonstrated in several experiments
\cite{BTS98}. This scheme allows us to coherently transfer
population between two suitable quantum states via a so-called
counterintuitive sequence of coherent light pulses that drive
transitions of a lambda or a multilevel system. It has some
advantages over the Landau-Zener method as the choice of control
parameters is easier and less constraints have to be fulfilled as we
do not eliminate the electronic states, which allows for faster
mapping times compared to the Landau-Zener method.

Note, that the main source of error, the decay of the cavity, is not
considered here. Up to now, the experimentally achieved cavity decay
rate $\gamma$ of a photonic crystal microcavity that couples to a
quantum dot is of the order of $\gamma\approx10^{10} \frac{1}{s}$
\cite{Taka07}. However, we do propose this scheme here, as cavity
decay rates might improve and the scheme might also be used in a
different setup.

For the system proposed in Section \ref{section1} the STIRAP method
is not as straight forward as for the setup we investigated in
\cite{SCG08} in Section V. The reason for this is, that after
elimination of the trion states in the system used so far, there are
two different couplings: $g_{a-}(S^+a^{\dagger}+h.c.)$ and
$g_{a+}(S^-a^{\dagger}+h.c.)$, with
$g_{a\pm}=\Omega_c\Omega_l/4\Delta_{T_{\pm}} $, where the first one
has to be made off-resonant: $g_{a-}\gg g_{a+}$, which means that
$\omega_h$ sets an upper limit to the coupling $g_{a+}$ (as the
condition for the adiabatic elimination is
$\Delta_{T_{\pm}}\gg\Omega_c,\Omega_l$). Therefore, we study the
STIRAP scheme for the system investigated in \cite{SCG08}, where
only the coupling $\propto(S^-a^{\dagger}+h.c.)$ is present.

In \cite{SCG08}, we study a singly charged QD where the electronic
states are Zeeman split by an external magnetic field in
growth/$z$-direction (Faraday geometry). The electronic state
$\ket{\uparrow}$ is coupled to the trion state $\ket{\Uparrow}$
(with angular momentum $+3/2$) by $\sigma^+$ circularly polarized
light and the electronic state $\ket{\downarrow}$ is coupled to the
trion state $\ket{\Downarrow}$ (with angular momentum $-3/2$) with
$\sigma^-$-polarized light. These transitions can be stimulated by a
$\sigma^+$-polarized cavity field and a $\sigma^-$-polarized
classical laser field, respectively. The trion states are mixed with
a resonant microwave field, whereas the electronic eigenstates are
unchanged as they are far detuned from the microwave frequency and
are now both coupled to the new trion eigenstates
$\ket{T_{\Uparrow}}=1/\sqrt{2}(\ket{\Uparrow}-\ket{\Downarrow})$ and
$\ket{T_{\Downarrow}}=1/\sqrt{2}(\ket{\Uparrow}+\ket{\Downarrow})$,
and form a double $\Lambda$ system (see \cite{SCG08} for a figure).

In a frame rotating with the laser frequency the Hamiltonian reads
\begin{eqnarray}\label{eq:opticalim2}
H&=&\frac{\Omega_c}{\sqrt{2}}\,(a^{\dagger}\,\ketbra{\downarrow}{T_{\Uparrow}}-a^{\dagger}\,\ketbra{\downarrow}{T_{\Downarrow}}+\text{h.c.})+
\frac{\Omega_l}{\sqrt{2}}(\ketbra{\uparrow}{T_{\Uparrow}}+\ketbra{\uparrow}{T_{\Downarrow}}+\text{h.c.})\nonumber\\
 &&+\delta'
a^{\dagger}a+\Delta_+\proj{T_{\Uparrow}}+\Delta_-\proj{T_{\Downarrow}}+
\tilde{\omega}_e S^z+H_{\text{hf}},
\end{eqnarray}
where $\delta'=\omega_c-\omega_l-\omega_{mw}$ and
$\Delta_{\pm}=\omega_{\Downarrow}-\omega_l\pm\Omega_{mw}$. Now, we
derive the Hamiltonian where only the trion has been eliminated. If
\begin{equation}
\Delta_{\pm}\gg\Omega_l,\Omega_c\sqrt{m}
\end{equation}
holds, the trion can be adiabatically eliminated. This leads to the
Hamiltonian
\begin{eqnarray}\label{eq:hamstirap}
H_{\mathrm{el}}=&g_a\left(S^+a+\text{h.c.}\right)+g_n\left(S^+b+\text{h.c.}\right)+\frac{A}{2N}S^zb^{\dagger}b+\delta'
a^{\dagger}a+ \tilde{\omega}_e S^z
\nonumber\\&+\left(\frac{\Omega_c^2}{2\Delta_-}+\frac{\Omega_c^2}{2\Delta_+}\right)
a^{\dagger}a\proj{\downarrow}
+\left(\frac{\Omega_l^2}{2\Delta_-}+\frac{\Omega_l^2}{2\Delta_+}\right)\proj{\uparrow}\nonumber\\
\end{eqnarray}
where the coupling
\begin{equation}
g_a=\frac{\Omega_c\Omega_l}{2}\left(\frac{1}{\Delta_+}-\frac{1}{\Delta_-}\right).
\end{equation}
We thus arrive, in addition to $H_{\mathrm{hf}}$, at an effective
Jaynes-Cummings-like coupling of the two electronic spin states to
the cavity mode governed by
\begin{equation}
  \label{eq:H1}
  g_a(S^+a+\text{h.c.}),
\end{equation}
i.e., the absorption of a cavity photon goes along with an upward
flip of the electron spin (and the emission of a photon into the
laser mode) and vice versa.

Next we present the STIRAP scheme: the couplings of the electronic
states $\ket{\uparrow}$ and $\ket{\downarrow}$ are given by the
optical fields ($g_a$) and the hyperfine coupling ($g_n$). The
Hamiltonian describing the system is given by (\ref{eq:hamstirap}).
It is blockdiagonal $H'=\bigoplus_m H_m$, where $m$ denotes the
initial photon number. The $(2m+1)$-dimensional Hamiltonian $H_m$,
describing the evolution of the "$m$-excitation subspace" can be
written in the Fock basis $\{\ket{m, \downarrow, 0},\ket{m-1,
\uparrow, 0}, \ket{m-1, \downarrow, 1},...\}$, where the first
number $m-k$ represents the Fock state of the cavity,
$\downarrow/\uparrow$ denotes the electron spin down/up state and
$k$ the excitation number of the nuclear spins. In this basis, $H_m$
reads:
\begin{eqnarray} \label{eqn:Heffmatrix}
H_{m}=\left(
  \begin{array}{cccccc}
  \Delta_{G_{0}} & g_{a}\sqrt{m} & 0& 0&0&\ldots\\
  g_{a}\sqrt{m} & \Delta_{E_{0}} & g_n\sqrt{1} &0&0&\ldots\\
  0& g_n\sqrt{1} & \Delta_{G1}& g_{a}\sqrt{m-1} &0&\ldots\\
  0 &0& g_{a}\sqrt{m-1} & \Delta_{E_{1}}&g_n\sqrt{2}&\ldots \\
  0& 0 & 0& g_n\sqrt{2} & \Delta_{G_{2}}&\ldots\\
  \vdots&\vdots&\vdots&\vdots&\vdots&\ddots
\end{array}\right),
\end{eqnarray}
with
$\Delta_{G_{k}}=\left(\delta'+\frac{\Omega_c^2}{2\Delta_-}+\frac{\Omega_c^2}{2\Delta_+}\right)(m-k)-\frac{A}{2N}k-\tilde{\omega}_e/2$,
and
$\Delta_{E_{l}}=\frac{\Omega_l^2}{2\Delta_-}+\frac{\Omega_l^2}{2\Delta_+}+\delta'(m-l-1)+\frac{A}{2N}l+\tilde{\omega}_e/2$
for $k\in\{0,1,2 .. m\}$ and $l\in\{0,1,2..m-1\}$. In the following,
we will denote the states with electron spin down as "ground
states", $\ket{G_k}=\ket{m-k, \downarrow, k}$ and the "excited
states" by $\ket{E_l}=\ket{m-l-1, \uparrow, l }$. The optical fields
couple states $G_k$ and $E_l$ with $k=l$ whereas states with $k=l+1$
are coupled by the hyperfine coupling (see figure
\ref{fig:stirap3}b).

We will show in the following, that by slowly increasing the laser
Rabi frequency and thus changing $g_a(t)$ such that $g_a(T)\gg g_n$,
at the final time $T$ (see figure \ref{fig:stirap3}a), an initial
state $\ket{\psi,\downarrow, 0}$ with no nuclear spin excitations in
the quantum dot and a state $\ket{\psi}$ in the cavity, evolves
under the adiabatic change of $H'$ to a state where the cavity is
empty and its state has been mapped to the nuclear spins:
\begin{equation}\label{eqn:mappingstirap}
\ket{\psi \downarrow 0}_{t=0}\rightarrow\ket{0 \downarrow
\psi}_{t\rightarrow T}
\end{equation}
for $T\rightarrow\infty$. A prerequisite for the mapping is, that
the ground states of the Hamiltonian are all degenerate within each
"$m$-excitation"-subspace so that we can keep track of the phases of
the individual eigenstates. This can be done by choosing the
parameters such that $\Delta_{G_{k}}$ does not depend on $k$, which
is fulfilled for
$\delta+\frac{\Omega_c^2}{2\Delta_-}+\frac{\Omega_c^2}{2\Delta_+}=-\frac{A}{2N}-\tilde{\omega}_e/2$
so that $\Delta_{G_{k}}=-\frac{A}{2N}m-\tilde{\omega}_e/2$
\footnote{It can be proven by induction, that $H_{m}$ has an
eigenvalue $E_{G_m}=\Delta_{G_m} \forall\, m $, i.e. that
$\det{(H_{m}-E_{G_{m}}\openone)}=0 \,\,\,\,\textrm{for all}\,\,\, m
 \in
\{0,1...\infty\}$ (where $m$ is the initial photon number).}. Hence,
the phases $\phi_{m}=\Delta_{G_{m}}\,t$ of the individual
eigenstates which the system acquires during the time evolution (for
perfect adiabaticity) are known and can be corrected, e.g., by
applying a magnetic field $-|B|\hat{z}$ for a time
$t=\frac{A/(2N)}{g_K \mu_K |B|}$ after the state transfer to the
nuclei. Here $g_K$ and $\mu_K$ denote the nuclear g-factor and the
nuclear magnetic moment, respectively.

\subsection{Numerical integration of the Schr\"{o}dinger equation}\label{sec:num}

To study the quality of the mapping of a state of the cavity to the
nuclear spins, we numerically integrate the Schr\"{o}dinger Equation
given by
\begin{equation}\label{eqn:sdg}
i\frac{\partial}{\partial t}\ket{\psi(t)}=H'(t)\ket{\psi(t)},
\end{equation}
where $H'$ is the Hamiltonian given by (\ref{eq:hamstirap}). The
simulation computes $\ket{\psi(t+Dt)}=e^{-iH'(t)Dt}\ket{\psi(t)}$ in
$T/Dt$ steps from $t=0$ to $t=T$. We assume, that the change of
$g_a(t)$ is quadratic in time, ensuring an initially slow and a
finally fast increase of $g_a(t)$ and $g_a(T)\gg g_n(T)$,
$g_a=10\,g_n\frac{t^2}{T^2}$. Parameters are chosen as follows: we
assume a hole g-factor $g_h=2.2$ and an electron g-factor
$g_e=0.48$; the number of nuclei $N=10^4$, the hyperfine coupling
constant $A=10\mu$eV, the laser and cavity Rabi frequency
$\Omega_c=\Omega_l=13\mu$eV, the detuning of the trion
$\omega_{\Downarrow}-\omega_l=103\mu$eV, the effective Zeeman
splitting $\tilde{\omega}_e=0\mu$eV and the microwave Rabi frequency
$\Omega_{mw}=50\mu eV$.
\begin{figure}
\begin{center}
\includegraphics{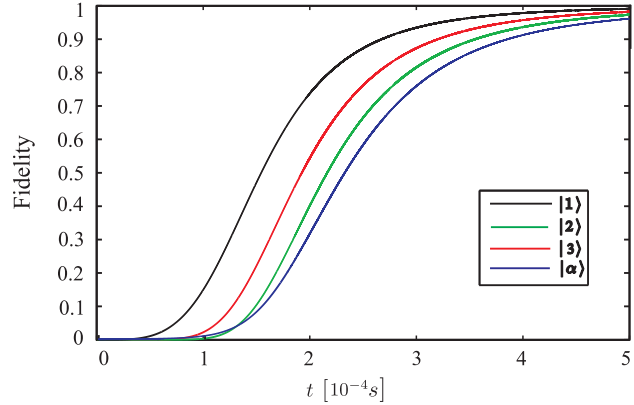}
\caption{Plot of the fidelity of a variety of states vs time during
the adiabatic evolution. $\ket{1}$, $\ket{2}$, $\ket{3}$ denote the
one, five, ten photon Fock states, respectively. $\ket{\alpha}$
denotes the coherent state with average photon number $5$. The total
evolution time is chosen to be
$T=5\cdot10^{-4}$s.\label{fig:fidelity3}}
\end{center}
\end{figure}
The fidelity of the mapping we are interested in is given by the
overlap of the numerically evolved state
$\rho(t)=\,\proj{\psi_{\text{si}}}$ and the ideal output
$\ket{\psi_{\text{id}}}$
\begin{equation}
F=\bra{\psi_{\textrm{id}}}\rho(t)\ket{\psi_{\textrm{id}}}.
\end{equation}
To achieve a fidelity close to one, the total evolution time is
chosen to be $T=5\cdot10^{-4}$s.
Figure \ref{fig:fidelity3} shows the fidelity plotted versus time
for
different kind of states that will be discussed in the following, illustrating the different aspects of mapping.\\

The one photon Fock state $\ket{1}$ is mapped in $T=5\cdot10^{-4}$s
with a fidelity of $F\approx0.99$ to the nuclear spins. To see that
not only population but also relative phases are properly mapped we
have simulated an approximately coherent state
$\ket{\alpha}=\exp{\left(-\frac{|\alpha|^2}{2}\right)}\sum_{k=0}^{20}\frac{\alpha^k}{\sqrt{k!}}\ket{k}$
with average photon number $|\alpha|^2=5$ and find a mapping
fidelity of $F\approx0.96$. Here the known phases $\phi_m$ have been
compensated.

\subsection{Error processes}\label{sec:errors}

The main error processes that lead to imperfections of the fidelity
are the "always-on" character of the hyperfine coupling, the
nonadiabaticity due to finite times, non-perfect polarization of the
nuclei and the decay of the cavity. These processes will be studied
in the following.

The fact that the \textit{hyperfine coupling is "always-on"} leads
to an "error" that is intrinsic to our system. Different to
conventional STIRAP that uses overlapping light pulses, we propose
to adiabatically increase the coupling $g_a(t)$ so that $g_a(T)\gg
g_n$ and therefore the mapping is imperfect as $g_n$ is constant and
can not be "switched off". Treating the coupling $g_n$ as a small
perturbation in first order perturbation theory at $t=T$, the
fidelity is found to be
\begin{equation}
F=\frac{|\bracket{\xi^0}{\xi}|^2}{\bracket{\xi}{\xi}}=\frac{1}{1+m\,g_n^2/{g_a(T)}^2}\approx
1-m\left(\frac{g_n}{g_a(T)}\right)^2,
\end{equation}
where $\ket{\xi^0}$ is the ideal output state for which $g_n=0$ (at
t=T),
and $\ket{\xi}=\ket{\xi^0}+\ket{\xi^1}+ ..$ is the unnormalized eigenstate of $H_m$.\\

Another error arises from the \textit{non-adiabaticity} due to
finite times of realistic processes. For a quantitative estimate of
the time $T$ that is needed for adiabatic passage to occur we use
the well-known adiabatic theorem \cite{Messiah2} and numerically
compute the minimum time $T$ fulfilling
\begin{equation}\label{eq:adth}
\frac{\abs{\bra{E_l^m(t)}\frac{\textrm{d}}{\textrm{dt}}(H'-\bigoplus_m
E_{G_m}\openone)\ket{\phi_0^m(t)}}}{\abs{E_l}^2}\leq
\delta_{\textrm{a}} \,\,\forall\, t\,\in [\,0 ,\, T\,].
\end{equation}
The left hand side of (\ref{eq:adth}) corresponds to the probability
to find the system in an excited state $\ket{E_l^m}$ different from
$\ket{\phi_0^m}$ (the (purely nuclear) eigenstates to the eigenvalue
$E=0$ of $H'-\bigoplus_m E_{G_m}\openone$) and the fidelity
decreases with $\delta_{\textrm{a}}$. For
$1/100<\delta_{\textrm{a}}<1/10$ the minimum time fulfilling
Equation (\ref{eq:adth}) for the mapping of one photon is in the
range of $1 \mu s>T>0.11 \mu s$.

To get an accurate description of the errors arising from
non-adiabaticity  we use a perturbative approach to treat
nonadiabatic corrections and compute the phases arising from
non-adiabaticity \cite{Shi04}. As the Hamiltonian $H'=\bigoplus_m
H_m$ is blockdiagonal, states with different initial photon numbers
$m$ do not couple so that we can treat every "$m$-photon" subspace
separately. Moreover, we can use nondegenerate perturbation theory
as the groundstates $\ket{G_k}$ that are degenerate within each
subspace are not coupled:
$\bra{G_{k'}}\frac{\textrm{d}}{\textrm{dt}}H\ket{G_k}$. Supposing to
be at $t=0$ in one of the groundstates $\ket{\phi_{G_k}(0)}$ of
$H_m'=H_{m}-E_{G_m}\openone$ slowly varying in time, the first order
correction of the energy eigenvalue $E_0=0$ of $H$ is given by
\begin{equation}
E_{m}^1=\sum_{l\neq k} \frac{\abs{\bra{\phi_{E_l}}
\frac{\textrm{d}}{\textrm{dt}}H_m'\ket{\phi_{G_k}}}^2}{E_l^3}.
\end{equation}
The phases $\phi_{m}^1=\int_0^T\frac{E_{m}^1}{\hbar}\,\textrm{dt}$
which the system acquires can be found by numerical integration of
$E_{m}^1$. For $T=5\cdot 10^{-4}$s and initial photon number $m=1$,
$\phi_{m}^1=-1.4\cdot 10^{-5}$ and for $m=2$, $\phi_{m}^1=0.004$,
respectively. Thus, as expected, the errors arising from
non-adiabaticity are small for sufficiently long times $T$.

\section{Quantum Interface in the bad cavity
limit}\label{sec:badcavity} In the previous Sections, we have shown
that a quantum interface can be achieved via direct mapping of the
cavity field to the nuclear spins of the QD.
\begin{figure}[b]
\begin{center}
\includegraphics{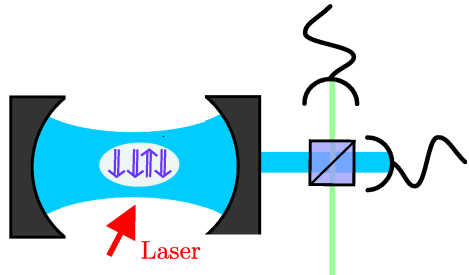}
\caption{Quantum teleportation can be used to write the state of a
traveling-wave light field onto the nuclei.}\label{fig:teleport3}
\end{center}
\end{figure}
\begin{figure}[t]
\begin{center}
\includegraphics{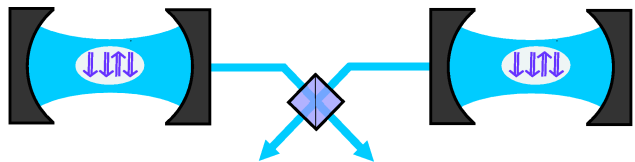}
\caption{Nuclear spins of quantum dots in two distant cavities can
be entangled by interfering the travelling wave output fields of the
two cavities at a beamsplitter and
measuring.}\label{fig:entanglenuc2}
\end{center}
\end{figure}
But we have also seen that the cavity lifetimes required for
high-fidelity storage are much larger than what is today's state of
the art, i.e., as $g_{1,2}\ll 1/\tau_\textrm{cavity}$, we are,
compared to the effective coupling, in the ``bad cavity limit''. A
second problem with this approach is that the quantum information we
want to map to the nuclei has to be coupled into a high-Q cavity.
This is notoriously difficult although theoretical proposals exist
\cite{CZKM97} that should avoid reflection completely. Both problems
can be circumvented employing ideas similar to \cite{CPGP03,KrCi04}
by using the two-mode squeezing Hamiltonian $H_\textrm{sq}$ [see
(\ref{eq:Hsq})] (note that we now return to the system proposed in
Section \ref{section1} for the rest of the paper). As discussed in
\cite{SCG08} and elaborated in more detail below, it is possible to
create entanglement between nuclei and the traveling-wave
\emph{output} field of the cavity. Then, quantum teleportation can
be used to write the state of another traveling-wave light field
onto the nuclei (figure \ref{fig:teleport3}) \footnote{This maps the
state up to a random (but known) displacement. It can be undone
using $H_{\textrm{bs}}$, where the cavity is pumped with strong
coherent light for a short time \cite{Par96}.}. This approach gives
an active role to cavity decay in the interface and can tolerate a
bad effective cavity as long as strong coupling is achieved in
(\ref{rotatingframeham}). Moreover, it does not require to couple
the quantum information into the cavity. Similarly $H_\textrm{bs}$
[(\ref{eq:Hbs})] enables read-out, by writing the state of the
nuclei to the output field of the cavity. The entanglement between
nuclear spins and output field can moreover be used to entangle
nuclear spins in two distant cavities by interfering the output
light of the cavities at a beamsplitter (figure
\ref{fig:entanglenuc2}).

\subsection{Entangling nuclei with the output
field}\label{sec:entangle} The Hamiltonian of the nuclear
spin-cavity system tuned to the squeezing interaction (\ref{eq:Hsq})
and coupled to the environment is given by
\begin{equation}\label{eq:hament}
H=g_2(a^\dag b^\dag + a b)+ia\int
\sqrt{\frac{\gamma}{2\pi}}c_{\omega}^{\dagger}\,d\omega+\text{h.c.}+\int\omega
c^{\dagger}_{\omega}c_{\omega}d\omega,
\end{equation}
where $c_{\omega}$ are the annihilation operators of the bath and
$\gamma$ the cavity decay constant. We have specialized
(\ref{eq:Hsq}) to the case $\omega_1=-\omega_2$ and transformed to
an interaction picture \footnote{As was already the case in
(\ref{eq:Hsq}) all optical operators are also taken in a fram
rotating with the laser frequency $\omega_l$.} with $H_0 =
\omega_1(a^\dag a-b^\dag b)+\omega_1\int
c^{\dagger}_{\omega}c_{\omega}d\omega$ and performed the
rotating-wave and Markov approximations in the description of the
cavity decay \cite{GZ00}. The quantum Langevin equations of cavity
and nuclear operators read \numparts
\begin{eqnarray}
  \dot{a}(t) &=& -i g_2\,
  b^{\dagger}(t)-\frac{\gamma}{2}a(t)-\sqrt{\gamma}c_{\textrm{in}}(t)\label{langevin1}\\
  \dot{b}(t) &=&-i g_2\, a^{\dagger}(t).\label{langevin2}
\end{eqnarray}
\endnumparts
Here, $c_\textrm{in}$ describes the vacuum noise coupled into the
cavity and satisfies
$[c_\textrm{in}(t),c_\textrm{in}^{\dagger}(t')]=\delta(t-t')$. The
solutions of (\ref{langevin1}) and (\ref{langevin2}) are given (for
$t\geq0$) by \numparts
\begin{eqnarray}
  a(t)&=&p_{-}(t)a(0)+q(t)b^{\dagger}(0)+\sqrt{\gamma}\int_0^t p_{-}(t-\tau)c_{\textrm{in}}(\tau)d\tau\label{eq:sol1a} \\
  b(t)&=& q(t)a^{\dagger}(0)+p_{+}(t)b(0)+\sqrt{\gamma}\int_0^t
  q(t-\tau)c_{\textrm{in}}^{\dagger}(\tau)d\tau
\label{eq:sol1b}
\end{eqnarray}
\endnumparts
where
\begin{eqnarray}\label{eq:alpha1}
  p_{\pm}=&e^{-\frac{1}{4}t\gamma }\left[\cosh{\left(\nu t\right)}\pm\frac{\gamma}{4\nu}\sinh{\left(\nu t\right)}\right],\\
q=&-i\frac{g_2}{\nu}e^{-\frac{1}{4}\gamma t}\sinh{\nu
t}\label{eq:alpha2},
\end{eqnarray}
with
\begin{equation}
\nu=\sqrt{\left(\frac{\gamma}{4}\right)^2+g_2^2}.
\end{equation}

While (\ref{eq:sol1a}) and (\ref{eq:sol1b}) describe a non-unitary
time-evolution of the open cavity-nuclei system, the overall
dynamics of system plus surrounding free field given by the
Hamiltonian in (\ref{eq:hament}) is unitary. Moreover, it is
Gaussian (see \ref{Appendix1}), since all involved Hamiltonians are
quadratic. Since all initial states are Gaussian (vacuum), the joint
state of cavity, nuclei, and output fields is a pure Gaussian state
at all times as well. This simplifies the analysis of the dynamics
and in particular the entanglement properties significantly:
The covariance matrix [defined by (\ref{eq:cov}) in \ref{Appendix1}]
of the system allows us to determine the entanglement of one part of
the system with another one. In particular, we are interested in the
entanglement properties of the nuclei with the output field.

The covariance matrix $\Gamma_{\textrm{ns-c-o}}$ of the pure
Gaussian state of nuclear spins, cavity and output field and thus
the covariance matrix $\Gamma_{\textrm{ns-o}}$ of the reduced
nuclei-output field system can be found by analyzing the covariance
matrix of the cavity-nuclei system $\Gamma_{\textrm{ns-c}}$.

The elements $\left<X\right>$ of the covariance matrix
$\Gamma_{\textrm{ns-c}}$ can be calculated by solving the Lindblad
equation evaluated for the expectation values $\left<X\right>$
\begin{equation}
\frac{d}{dt}\left<X\right>=i\left<
[H_{\textrm{sq}},X]\right>+\frac{\gamma}{2}\left(\left<2a^{\dagger}Xa\right>-\left<Xa^{\dagger}a\right>-\left<a^{\dagger}aX\right>\right).
\end{equation}
We thus find the covariance matrix of the cavity-nuclei system to be
\begin{equation}\label{eq:A}
\Gamma_{\textrm{ns-c}}= \left(\begin{array}{cccccccc}
m&0&0&k\\
0&m&k&0\\
0&k&n&0\\
k&0&0&n
\end{array}\right),
\end{equation}
where
\numparts
\begin{eqnarray}
  m&=&e^{-\frac{\gamma
t}{2}}\left[\frac{\gamma}{\nu}\sinh{\left(2\nu t\right)}\right.
\left.+\left(\frac{g_2^2}{\nu^2}+\frac{\gamma^2}{8\nu^2}\right)\cosh{\left(2\nu t\right)}+\frac{g_2^2}{\nu^2}\right]-1,\label{eq:mnk1}\\
n&=&1+32\frac{g_2^2}{\nu^2}e^{-\frac{\gamma t}{2}}\sinh{\left(\nu
t\right)}^2,\label{eq:mnk2}\\
 k&=&e^{-\frac{\gamma t}{2}}
\left[\frac{g_2\gamma}{2\nu^2} \sinh{\left(\nu
t\right)}^2+\frac{g_2}{\nu}\sinh{\left(2\nu t\right)}\right].
\label{eq:mnk3}
\end{eqnarray}
\endnumparts
According to \cite{Wil36} there exists a symplectic transformation
$S$ (cf. \ref{Appendix1}) such that
$\Gamma_{\textrm{D}}=S\Gamma_{\textrm{ns-c}}S^T=\textrm{diag}(\lambda^s_1,\lambda^s_1,\lambda^s_2,\lambda^s_2)$
where $\{\lambda^s_1,\lambda^s_2\}$ are the symplectic eigenvalues
of $\Gamma_{\textrm{ns-c}}$. This allows us to calculate the
covariance matrix of the pure nuclei-cavity-output field system
\begin{equation}
\Gamma_{\textrm{ns-c-o}}=S'\Gamma_{\textrm{D'}}(S'^{-1})^T
\end{equation}
with $\Gamma_{\textrm{D'}}$ in $2\times2$ block-matrix form
\begin{equation}\label{eq:gescavma}
\Gamma_{\textrm{D'}}=\left(\begin{array}{cccc}
\cosh{(2r_1)}\openone_{2}& &\sinh{(2r_1)}\sigma_z& \\
 &\cosh{(2r_2)}\openone_{2}& &\sinh{(2r_2)}\sigma_z\\
\sinh{(2r_1)}\sigma_z& &\cosh{(2r_1)}\openone_{2}& \\
&\sinh{(2r_2)}\sigma_z&&\cosh{(2r_2)}\openone_{2}
\end{array}\right),
\end{equation}
where $\cosh{r_1}=\lambda^s_1$ and $\cosh{r_2}=\lambda^s_2$ and
$S'=\left(\begin{array}{cc} S& \\
 &\openone_{4\times4}
\end{array}\right)$.
One of the symplectic eigenvalues $\{\lambda^s_1,\lambda^s_2\}$  is
1, indicating a pure - and therefore unentangled - mode in the
system. That implies that there is a \emph{single} ``output mode''
in the out-field of the cavity to which the cavity-nuclear--system
is entangled and we can thus trace out the unentangled output mode.

The procedure for entangling the nuclei with the output field
(write-in) is: let $H_\text{sq}$ act for time $t_1$ to create a
two-mode squeezed state $\psi(g_2,t_1)$: nuclei entangled with
cavity and output field. To obtain a state in which the nuclei are
only entangled to the output field, we switch the driving laser off
$(g_2=0)$ and let the cavity decay for a time $t_2\gg
\tau_\text{cav}$, obtaining an almost pure two-mode squeezed state
of nuclei and the output mode. We define the coupling as
\begin{equation}
g_t= \cases{g_2, t<t_1}{0,  t\geq t_1}
\end{equation} For the parameters used in Section \ref{sec:coupling},
$g_{2}\sim1.9\cdot10^{-3}\mu$eV.
\begin{figure}
\begin{center}
\includegraphics{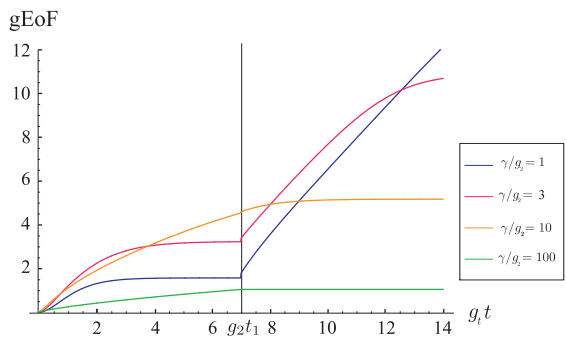}
\caption{Plot of the Gaussian entanglement of formation (gEoF) of
the nuclei with the output field vs $t$ for different values of
$\gamma/g$. At $g_2t_1=7$ the coupling is switched off. The curve
saturates when all excitations have leaked out of the
cavity.}\label{fig:plotent1}
\end{center}
\end{figure}
\begin{figure}
  \centering
  \subfigure[][]{
    \label{ee1}
    \includegraphics[width=0.4\textwidth]
{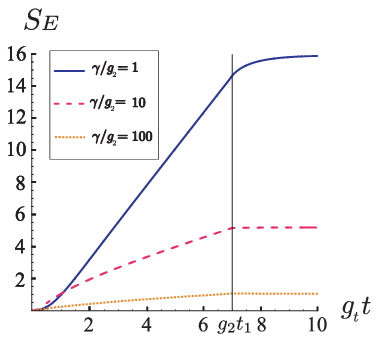}}\hfill
  \subfigure[][]{
    \label{ee2}
    \includegraphics[width=0.4\textwidth]{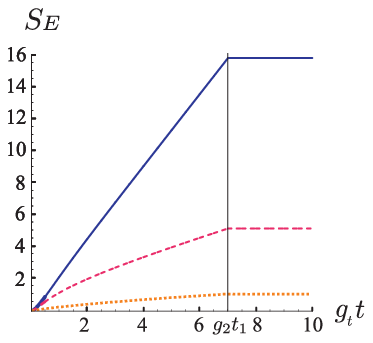}
  }
  \caption{a) Plot of the entanglement entropy $S_E$ of the
nuclei+cavity with the output field vs $t$ for different values of
$\gamma/g$. At $g_2t_1=7$ the coupling is switched off. b) Plot of
the entanglement entropy $S_E$ of the nuclei with the cavity+output
field vs $g_2t$ for different values of
$\gamma/g_2$.}\label{fig:plotentkombi2}
\end{figure}
The entanglement of the different subsystems can be quantified: We
compute the Gaussian entanglement of formation (gEoF) \cite{WGK+03}
of the reduced covariance matrix of the nuclei-output field--system
to quantify the entanglement of the nuclei with the output field
(see figure \ref{fig:plotent1}). The gEoF measures how costly it is
to generate a state by mixing pure Gaussian states. It gives an
upper bound to the entanglement of formation (EoF) and is in the
present case equivalent to the logarithmic negativity \cite{Ple05}.
The entanglement of the pure cavity-nuclei-output mode--system can
be quantified using the entanglement entropy $S_E$ \cite{BBPS96}. We
plot $S_E$ for the nuclei-cavity system with the output mode [see
figure \ref{fig:plotentkombi2}a)] and of the nuclei with the
cavity-output mode--system [see figure \ref{fig:plotentkombi2}b)].
The entanglement is plotted versus $g_2t$ for different ratios of
the cavity decay constants and the coupling, $\gamma/g_2$.

\subsection{Write-in: Teleportation channel}
The entangled state between nuclei and the cavity output field
allows us to map a state of a traveling light field to the nuclei
using teleportation (see figure \ref{fig:teleport3}) \cite{BrKi98}.

To realize the teleportation, a Bell measurement has to be performed
on the output mode of the cavity and the signal state to be
teleported. This is achieved by sending the two states through a
50:50 beam splitter and measuring the output quadratures
\cite{BrKi98}. To be able to do this, we need to know $B_0$, the
output mode of the cavity. In the following, we derive an exact
expression for this mode.

We fix a time $t$ and denote by $B(y,t)$, $y \in \mathbb{N}$ a
complete set of bath modes outside the cavity. $B(y,t)$ can be
expressed as a superposition of bath operators $c(x,t)$
\begin{equation}\label{outputmode}
B(y,t) = \int z(y,x,t)c(x,t)dx
\end{equation}
where we introduce a complete set of orthonormal mode functions
$z(y,x,t)$. The bath operators $c(x,t)$ are known from the
input-output relations \cite{GZ00}
\begin{equation}\label{eqn:inputoutput}
c(x,t)=\frac{\sqrt{\gamma}}{2}a(t-x)\chi_{[0,t]}(x),
\end{equation}
where $a(t)$ is given by (\ref{eq:sol1a}). To calculate $B(y,t)$ we
thus need to determine $z(y,x,t)$. This can be done, calculating the
variance
$\left<c^{\dagger}(x,t),c(x',t)\right>=\left<c^{\dagger}(x,t)c(x',t)\right>-\left<c^{\dagger}(x,t)\right>\left<c(x',t)\right>$
following two different pathways: With (\ref{eqn:inputoutput}) we
find
\begin{equation}\label{variance1}
\left<c^{\dagger}(x,t),c(x',t)\right>=\frac{\gamma}{4}q(t-x')q(t-x)^*,
\end{equation}
where $q(t)$ is given by (\ref{eq:alpha2}). Another way to express
$c(x,t)$ follows from (\ref{outputmode}):
\begin{equation}\label{outputmode2}
c(x,t) = \sum_y z(y,x,t)^*B(y,t).
\end{equation}
As shown in Section \ref{sec:entangle} there exists only one output
mode which we label $y=0$. This mode contains all the output
photons. Therefore
$\left<B(y,t)^{\dagger}B(y',t)\right>=K\,\delta_{y0}\delta_{y'0}$
and the variance using (\ref{outputmode2}) reads
\begin{equation}\label{variance2}
\left<c^{\dagger}(x,t),c(x',t)\right>= K\, z(0,x,t)z(0,x',t)^*.
\end{equation}
Comparing (\ref{variance1}) to (\ref{variance2}) we find
\begin{equation}
z(0,x,t)=\frac{q(t-x)^*}{\sqrt{\int |q(t-x)|^2 dx}}
\end{equation}
and $K=\frac{\gamma}{4}\sqrt{\int |q(t-x)|^2 dx}$ and we have thus
fully determined $B(0,t)$ (see figure \ref{fig:outputmode}). Note
that the bath modes are given in a frame rotating with
$\omega_1+\omega_l$ to which we transformed in Section
\ref{section1} ($\omega_l$) and Section \ref{sec:badcavity}
($\omega_1$).

Therefore a state of a traveling light field can be teleported to
the nuclear spins up to a random displacement that arises from the
teleportation protocol \cite{Vai94, BrKi98}. The random displacement
can be undone, letting the beam-splitter interaction $H_{bs}$ [given
by (\ref{eq:Hbs})] act for a short time, while pumping the cavity
with intense coherent light as suggested in \cite{Par96}.

Next, we want to consider the quality of the teleportation. Whereas
before (see figures \ref{fig:plotent1} and \ref{fig:plotentkombi2})
the time evolution of the system for a fixed switch-off time $
g_2t_1=7$ was considered, we now consider the "final" entangled
state of nuclei and output field depending on $g_2t_1$, where the
cavity has decayed to the vacuum state while the nuclei are (still)
stationary.

The fidelity with which a quantum state can be teleported onto the
nuclei is a monotonic function of the two-mode squeezing parameter
\begin{equation}
r_1=\frac{1}{2}\textrm{arccosh(\textit{$m(t=t_1)$})}
\end{equation}
 with $m$ defined in (\ref{eq:mnk1}). A typical benchmark \cite{HWPC05} is the
average fidelity with which an arbitrary coherent state can be
mapped. This fidelity has a simple dependence on the two-mode
squeezing parameter $r_1$ of the state used for teleportation and is
given by \cite{Fiu02}
\begin{equation}
F_{\textrm{tel}}=\frac{1}{1+e^{-2r_1}}.
\end{equation}
We plot the teleportation fidelity dependent on the switch-off time
\begin{figure}
\begin{center}
\includegraphics[scale=1.5]{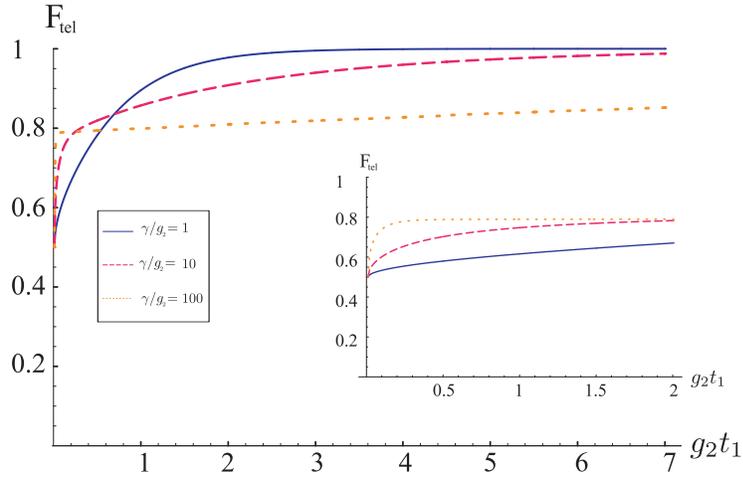}
\caption{Plot of the teleportation fidelity vs $g_2t_{1}$ for
different values of $\gamma/g_2$.}\label{fig:plotsfidelityent}
\end{center}
\end{figure}
$t_1$ (see figure \ref{fig:plotsfidelityent}).

Already for $r_1(t_{1})\sim1$ fidelities above $0.8$ are obtained.
After switching off the coupling we have to wait for the cavity to
decay which typically happens on a nanosecond timescale and does not
noticeably prolong the protocol.

\subsection{Read-out}
The beamsplitter Hamiltonian $H_\textrm{bs}$ [given by
(\ref{eq:Hbs})] enables read-out of the state of the nuclei by
writing it to the output field of the cavity. The quantum Langevin
equations of cavity and nuclear operators lead to almost identical
solutions as for $H_\textrm{sq}$ [see (\ref{eq:sol1a}) and
(\ref{eq:sol1b})]: of course, now $a(t)$ is coupled to $b(t)$
instead of $b^\dag(t)$ but the only other change to (\ref{eq:sol1a})
and (\ref{eq:sol1b}) is to replace $\nu$ by
\begin{equation}
\tilde\nu=\sqrt{(\gamma/4)^2-g_1^2}.
\end{equation}
This has the effect that all terms in (\ref{eq:sol1a})and
(\ref{eq:sol1b}) show exponential decay with $t$. The decay of the
slowest terms $\sim e^{-2\frac{g_1^2}{\gamma}t}$ sets the timescale
for read-out. To calculate the read-out fidelity, we need to know
the state of the output field at time $t=T$. We assume that the
state we want to read-out is a coherent state with displacement
$\alpha_{\textrm{ns}}$ at time $t=0$ fully described by its
covariance matrix $\gamma_b(0)=\openone$ and its displacement
$d_{b}(0)=\left<b\right>=\alpha_{\textrm{ns}}$ (while cavity and
output field are in the vacuum state at $t=0$). As the norm of the
displacement $\|d(t)\|$ of the nuclei-cavity-output system
\begin{equation}
d(t)=\left(\begin{array}{c}
d_a(t)\\
d_b(t)\\
d_{B_0}(t)
\end{array}\right)=\left(\begin{array}{c}
\left<a(t)\right>\\
\left<b(t)\right>\\
\left<B_0(t)\right>
\end{array}\right)
\end{equation}
 does not change under the beamsplitter transformation, the
displacement of the output mode $B_0$ is given by
\begin{eqnarray}
|d_{B_0}(t)|&=\sqrt{\|d(0)\|^2-d_{a}(t)^2-d_{b}(t)^2}\nonumber\\&=\sqrt{1-(|q(t)|^2+|p_{+}(t)|^2)}|\alpha_{\textrm{ns}}|
\end{eqnarray}
where $q(t)$ and $p_{+}(t)$ are defined by (\ref{eq:sol1a}) and
(\ref{eq:sol1b}) with $\nu$ replaced by $\tilde\nu$.

\begin{figure}
\begin{center}
\includegraphics[scale=1.25]{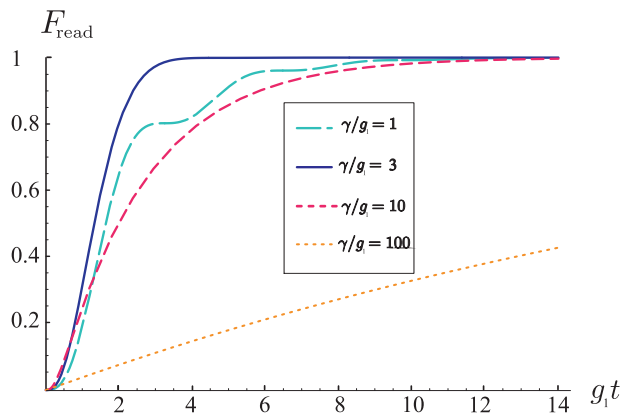}
\caption{Plot of the read-out fidelity vs $ g_1t$ for different
values of $\gamma/g_1$}\label{fig:plotsfidelityread}
\end{center}
\end{figure}

At finite times, the nuclear excitations and the cavity have not
fully decayed which leads to a loss of amplitude of the mapped
state. The loss is very small for sufficiently large $T$. To assure
high fidelity even for states with large photon number, we can
amplify the output field as in Section \ref{sec:lzfock}. Then the
state of the output field is $(\gamma_{B_0}, d_{B_0})=(\kappa
\openone, \alpha_{\textrm{ns}})$ with $\kappa$ as defined in Section
\ref{sec:lzfock}. This leads to a read-out fidelity (see
figure~\ref{fig:plotsfidelityread}) given by
\begin{eqnarray}\label{eq:readoutfid}
F_{\text{read}}=&\abs{\bracket{\openone,\alpha_{\textrm{ns}}}{\,
\kappa\openone,\alpha_{\textrm{ns}}}}^2\nonumber=1-(|q|^2+|p_{+}|^2),
\end{eqnarray}
where we have used relations for the transition amplitudes (as in
Section \ref{sec:lzfock}) given by \cite{Scu98}.

\subsection{Output mode}
\begin{figure}[b]
  \centering
  \subfigure[][]{
    \label{outmode1}
    \includegraphics[width=0.48\textwidth]
{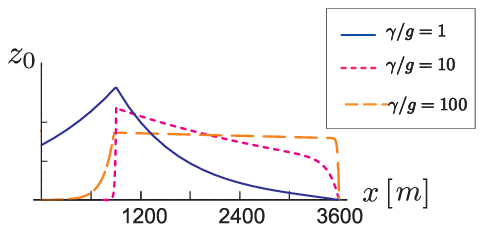}}\hfill
  \subfigure[][]{
    \label{outmode2}
    \includegraphics[width=0.46\textwidth]{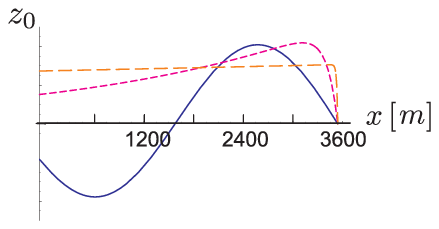}
  }
  \caption{The output mode in one dimension: Plot of $z_0$ vs position
$x$, where $x=0$ is the position of the cavity a) Write-in: The
squeezing interaction is "on" for $3\mu$s and then switched off. b)
Read-out: For $\gamma\gg g_2$ the excitations do not have fully
decayed to the output mode after $t=12 \mu$s. The read-out fidelity
given by (\ref{eq:readoutfid}) corresponds to the probability that
the excitations in the nuclear spins have decayed into the output
mode of the cavity. For $\gamma/g_1=1$ and $\gamma/g_1=10$ the
read-out fidelity is $F_{\text{read}}>0.98$ after $t\approx16-20
\mu$s. For $\gamma/g_1=100$ however, it takes $\approx200\mu$s to
achieve $F_{\text{read}}>0.98$. Note that for input and output modes
that have similar shapes (e.g. for a network), it is best to
consider the case where $\gamma/g_1\gg1$.)}
 \label{fig:outputmode}
\end{figure}

In figure \ref{fig:outputmode} we plot the output mode of the cavity
given by (\ref{outputmode}) for write-in and read-out, respectively,
and for several choices of the parameters $g_{1,2}$ and $\gamma$. We
are considering here only the idealized case of a one-sided and
one-dimensional cavity. In general, the actual geometry of the
cavity at hand has to be taken into account to determine $B_0$. In
the following we briefly discuss the shape of the mode-function. It
provides some insight into the dynamics of the mapping process,
since due to (\ref{eqn:inputoutput}) the weight of $c(x,T)$ in
$B(0,t)$ reflects the state of the cavity mode at time $t-x$ in the
past.

\emph{Write-in: } Let us consider the two extreme cases of very
strong and very weak cavity decay. In the former case ($\gamma\gg
g_2$) the cavity mode can be eliminated, i.e., the nuclear spins
couple directly and with constant strength $\sim g_2^2/\gamma$ to
the output field: $z_0$ is a stepfunction which is $0$ for $g_2=0$
and constant otherwise. This is reflected in figure
\ref{fig:outputmode}, where for $\gamma=100 g_2$ most of the
excitations decay directly to the outputmode such that $z_0$ takes a
"large" value at the time the squeezing is switched on and then
increases only slowly in time. After switching the squeezing
interaction off the cavity quickly decays to the vacuum. For
$\gamma\ll g_2$, instead, two-mode squeezing builds up in the
nuclei--cavity system as long as the squeezing interaction is on
($3\mu$s in figure \ref{fig:outputmode}) and after $g_2$ is switched
off the cavity decays to its standard exponential output mode. The
intermediate cases in figure~\ref{fig:outputmode}a) show the
shifting weight between ``initial step-function'' and subsequent
exponential decay.

\emph{Read-out: } In the case of the beamsplitter interaction, the
same cases can be distinguished. For large $\gamma/g_1$, the cavity
can be eliminated and the nuclear spins are mapped directly to the
exponential output mode of a cavity decaying with an effective rate
$g_1^2/\gamma$. For smaller $\gamma$, the output mode reflects the
damped free evolution of the nuclei--cavity system, which in this
case includes oscillations (excitations are mapped back and forth
between nuclei and cavity at rate $g_1$) in absolute value and
phase.
\subsection{Linear Optics with the Nuclear Spin Mode}
The interaction we have described can not only be used to \emph{map}
states to the nuclear spin ensemble but also for \emph{state
generation} and transformation. In fact, from a nuclear spin mode in
the vacuum state, all single mode Gaussian states can be prepared.
To see this, we have to show how any desired $2\times2$ correlation
matrix $\Gamma$ and displacement $d\in\mathbbm{C}$ can be obtained.

As we remarked already when discussing the write-in via
teleportation, the beam-splitter Hamiltonian $H_\text{bs}$ can be
used to realize displacements of the nuclear mode.  Driving the
cavity mode with a strong laser to a coherent state with amplitude
$\alpha$ (and the same phase as $d$) and switching on $H_\text{bs}$
for a time $t=|\beta|/(g_1|\alpha|)$ provides in the limit of large
$\alpha$ a good approximation to the displacement operation by
$\beta$ \cite{Par96}.

Concerning the CM, we use that every CM of a pure Gaussian state is
of the form $\Gamma=ODO^T$, where $D$ is a positive diagonal matrix
with determinant one and $O$ is orthogonal and symplectic. $O$ can
be seen as the effect of time evolution under some quadratic
Hamiltonian acting on the single-mode squeezed state with CM $D$. In
the single mode case, any $O$ represents a phase shift and is
obtained by letting the nuclear system evolve ``freely'' (without
laser coupling, i.e. a polarized electron interacts off-resonantly
with the nuclei) according to the Hamiltonian $\propto b^\dag b$ for
some time. Thus the state with CM $\Gamma$ can be generated in a
two-step process: first generate the state with $\Gamma=D$, then
apply $O$.

While in the preceding paragraphs we could show how to realize
\emph{operations} that can act on any input state, no such
possibility seems to exist for squeezing in our context. Instead we
show how to obtain the pure single mode squeezed state with CM $D$
from the vacuum state. Letting $H_\text{sq}$ act on the vacuum
results in a two-mode squeezed state with squeezing parameter $r_2$.
Performing a homodyne measurement (of the $X$ quadrature) on the
optical part of this state projects the nuclear system into a
squeezed state with squeezing $r_1= \ln[\cosh(2r_2)]/2$
\cite{GiCi02}, thus given enough two-mode squeezing, any CM $D$ can
be produced.

One can go even further and \emph{simulate} evolution according to
any quadratic Hamiltonian on the nuclear-optical system: According
to \cite{KHGC02}, the Hamiltonian given by (\ref{eq:Heff1b}) with
the interaction part $g_1 ab^{\dagger} + g_2 a^{\dagger} b^{\dagger}
+ \textrm{h.c.}$ enables simulation of any Hamiltonian quadratic in
$a,b,a^\dag,b^\dag$.

\section{Remarks on internal nuclear dynamics and approximations} \label{sec:approx}
With regard to the realization of the proposed protocol and the
applicability of the approximations leading to the Hamiltonians
(\ref{eq:Hbs}), (\ref{eq:Hsq}) there are three aspects to consider:
spontaneous emission of the quantum dot, the internal nuclear
dynamics and errors in the bosonic description. We assume the strong
coupling limit of cavity-QED and neglect spontaneous emission of the
quantum dot. The other two aspects will be studied in the following.
Note that the results on the internal nuclear dynamics are
corroborated by independent work of Kurucz \emph{et al.}
\cite{KST+Fl09}. They introduce the bosonic description to analyze
the performance of a nuclear spin quantum \emph{memory} and show
that the performance of the memory is enhanced due to a detuning
between excitations in the mode $b$ versus those in other modes
$b_{k\not=0}$ and that secular dipolar terms do not affect the
memory.
\subsection{Internal nuclear dynamics}\label{sec:nucinter}
Up to now, we have focused exclusively on the hyperfine interaction
and neglected ``internal'' nuclear dynamics, dominated by dipolar
and quadrupolar interactions. Moreover, the hyperfine coupling leads
to a dipolar interaction between nuclei mediated by the electron. We
study the dipolar interaction between nuclear spins which is
significantly weaker than $g_n$, $g_1$ and $g_2$: the energy scale
for dipolar interaction between two nuclei has been estimated
$\sim10^{-5}\mu$eV for GaAs \cite{SKL03}. However, since for
$10^4-10^6$ nuclei there are many of these terms, they might play a
role at the $10-50\mu$s time scales considered.

\subsubsection{Dipolar interaction}\label{sec:dipoledipole}
The Hamiltonian of the direct dipolar interaction between $N$ nuclei
is given by \cite{Sli90}
\begin{equation}
H_{dd}=-\frac{\mu_0}{4\pi}\frac{1}{2}\sum_{i=1}^N\sum_{j\neq
i=1}^N\frac{\mu_i\mu_j}{I_{i}I_{j}}\frac{1}{r_{ij}^3}\left(\frac{3(\mathbf{I}_i\mathbf{r}_{ij})(\mathbf{I}_j\mathbf{r}_{ij})}{r_{ij}^2}-\mathbf{I_i}\mathbf{I_j}\right),
\end{equation} where $\mathbf{r}_{ij}$ is the vector connecting spins
$i$ and $j$ and $\bm\mu_i=(\mu_i/I_i)\mathbf{I}_i$ is the magnetic
moment of the nuclear spin operator $\mathbf{I}_i$. $H_{dd}$ can be
written as
\begin{eqnarray}\label{eqn:hdd}
H_{dd}=&\sum_{i=1}^N\sum_{j\neq
i=1}^N\tilde{\gamma}_{ij}[A_{ij}I_i^zI_j^z+B_{ij}I_i^+I_j^-+(C_{ij}I_i^+I_j^++D_{ij}I_i^zI_j^-+\textrm{h.c.})]
\end{eqnarray}
where $A_{ij}=1-3\cos^2{\theta_{ij}}$,
$B_{ij}=-\frac{1}{2}(1-3\cos^2{\theta_{ij}})$,
$C_{ij}=-\frac{3}{4}\sin^2{\theta_{ij}}e^{-2i\phi_{ij}}$,
$D_{ij}=-\frac{3}{2}\sin{\theta_{ij}}\cos{\theta_{ij}}e^{i\phi_{ij}}$
and $\tilde{\gamma}_{ij}=\mu_0\mu_i\mu_j/4\pi r_{ij}^3$. In GaAs the
nearest-neighbor dipolar interaction strength is around
$\tilde{\gamma}=10^{-5}\mu$eV \cite{SKL03}. We want to calculate the
strength of the dipolar interaction between the main bosonic mode
(that is defined as the mode that is coupled to the electron spin)
and other bath modes (here, we no longer assume homogeneous coupling
of the nuclei to the electron). We therefore write the Hamiltonian
in terms of collective nuclear spin operators, use, in a next step,
the bosonic approximation and finally separate the relevant terms
(the ones which couple the main bosonic mode to bath modes) and
calculate the coupling strength of the main mode to the bath modes.

For highly polarized nuclear spins, the first term of $H_{dd}$ can
be written as
\begin{equation}
\sum_{i=1}^N\sum_{j\neq
i=1}^N\tilde{\gamma}_{ij}A_{ij}I_i^zI_j^z\approx\frac{1}{2}\sum_{i=1}^N\sum_{j\neq
i=1}^N
\tilde{\gamma}_{ij}A_{ij}\left(\frac{1}{2}-I_i^+I_i^--I_j^+I_j^-\right),\nonumber
\end{equation}
where we write $I_i^z=-1/2+I_i^+I_i^-$ and neglect the second order
term $I_i^+I_i^-I_j^+I_j^-$ which requires two excitations to be
non-zero; thus in the highly polarized case the contribution from
these terms is by a factor of $p=(1-P)/2$ smaller than the terms we
keep. The last term is (for spin 1/2-nuclei)
\begin{equation}
\sum_{i=1}^N\sum_{j\neq
i=1}^N\tilde{\gamma}_{ij}D_{ij}I_i^zI_j^-\approx-\frac{1}{2}\sum_{i=1}^N\sum_{j\neq
i=1}^N\tilde{\gamma}_{ij} D_{ij}I_j^-,\nonumber
\end{equation}
neglecting higher order terms.
In extension to the definition of the collective operators $A^{\pm}$
in Section \ref{section1}, which we now label $A^{\pm}_0$, we
introduce a complete set of collective operators
$A^-_k=\sum_i\alpha_i^{(k)}I_i^-$ with $k=0,..,N-1$ with an
orthogonal set of coefficients $\alpha_i^{(k)}$ for which $\sum_i
\alpha_i^{k}=1$ for every collective mode $k$. Defining a unitary
matrix U with columns
${\bm\alpha^{(k)}}=\left(1/\sqrt{\sum_i\alpha_i^{(k)2}}\right)\left(\alpha^{(k)}_1,..,\alpha^{(k)}_N\right)^T$
we can write
\begin{equation}
(I_1^-,..,I_N^-)^T=U\mathbf{A}^-
\end{equation}
where
$\mathbf{A}^-=\mathrm{diag}\left(\frac{1}{\sqrt{\sum_i{\alpha_i^{(0)}}^2}},..,\frac{1}{\sqrt{\sum_i{\alpha_i^{(N-1)}}^2}}\right)(A_0^-,..,A_{N-1}^-)^T$.
Writing $H_{dd}$ in terms of the
collective operators $A^{-,+}_k$ and neglecting higher order terms,
\begin{equation}\label{eqn:hdd1}
H_{dd}=\mathbf{A}^+U^{\dagger}SU\mathbf{A}^-+(\mathbf{A}^-U^{\dagger}MU\mathbf{A}^--\frac{1}{2}DU\mathbf{A}^-+\textrm{h.c.}).
\end{equation}
Here, $M_{ij}=\tilde{\gamma}_{ij}C_{ij}$ for $i\neq j$, $M_{ij}=0$
for $i=j$, $S_{ij}=\tilde{\gamma}_{ij}B_{ij}$ for $i\neq j$ and
$S_{ii}=\sum_{l=1}^N \tilde{\gamma}_{il}A_{il}$ for $i=j$. $D$ is a
vector with entries $D_j=\sum_{i\neq
j=1}^N\tilde{\gamma}_{ij}D_{ij}$.
Next, we write $H_{dd}$ in terms of bosonic operators, using the
bosonic approximation introduced in Section \ref{section1}, and map
$\mathbf{A}^-\longrightarrow\mathbf{b}=\left( b_{0},..,
b_{N-1}\right)^T$. This allows to separate relevant terms of
$H_{dd}$, which couple the main bosonic mode $b_{0}$ to other (bath)
modes $b_{k}$. Isolating the terms containing $b_0$, we find
\begin{eqnarray}\label{eq:dipoledipole}
\fl &b_{0}\left[\sum_{k\neq0}(U^{\dagger}2MU)_{0k}b_{k}
+(U^{\dagger}S U)_{0k}b^{\dagger}_k-\frac{1}{2}D_kU_{0k}\right]
+\textrm{h.c.}\nonumber\\&+
(U^{\dagger}SU)_{00}b^{\dagger}_0b_{0}+(U^{\dagger}2MU)_{00}b_0b_{0}+\textrm{h.c.}.
\end{eqnarray}
where the notation $(U^{\dagger}S U)_{0l}$ denotes the element
$(0,l)$ of the matrix $U^{\dagger}S U$. The first term describes the
passive coupling of the main mode $b_0$ to other modes $b_k$ and
acquires a factor of two as the terms that describe the active
coupling in (\ref{eq:dipoledipole}) can be written
$b_{l}(U^{\dagger}MU)_{lk}b_{k}+b_{k}(U^{\dagger}MU)_{kl}b_{l}=b_{l}(U^{\dagger}2MU)_{lk}b_{k}$
as $(U^{\dagger}MU)_{kl}=(U^{\dagger}MU)_{lk}$: the entries of $U$
are real so that $(U^{\dagger})^T=U$ and $M=M^T$, i.e.,
$M_{ij}=-\tilde{\gamma}_{ij}\frac{3}{4}\sin^2{\theta_{ij}}e^{-2i\phi_{ij}}=M_{ji}$
as $\phi_{ji}=\pi+\phi_{ij}$. The second term in
(\ref{eq:dipoledipole}) describes the passive coupling of $b_0$ to
the modes $b_k^{\dagger}$ and the third term displaces the main
mode. The last two terms describe a constant energy shift ($\sim
b_0^{\dagger}b_0$) and a squeezing term ($\sim
b_0b_0+\textrm{h.c.}$), respectively.

The terms that couple the main mode $b_{0}$ to bath modes can be
written as
\begin{eqnarray}\label{eq:coup}
&b_{0}\left(\sum_{k\neq0}(U^{\dagger}2MU)_{0k}b_k +(U^{\dagger}S
U)_{0k}b^{
\dagger}_k-\frac{1}{2}D_kU_{0k}\right)+\textrm{h.c.}\nonumber\\&=b_{0}\left(c_1\tilde{b}_1+c_2\tilde{b}^{\dagger}_2-\frac{1}{2}\sum_{k\neq0}D_{k}U_{0k}\right)+\textrm{h.c.}
\end{eqnarray}
where the linear combinations of bosonic modes $b_k$,
$b^{\dagger}_k$ can be transformed to bosonic modes $\tilde{b}_1$
and $\tilde{b}^{\dagger}_2$. The coupling strength of $b_{0}$ to the
first term in (\ref{eq:coup}) is given by
\begin{eqnarray}
[c_1\tilde{b}_1,(c_1\tilde{b}_1)^{\dagger}]&=|c_1|^2=\sum_{k\neq0}|(U^{\dagger}2MU)_{0k}|^2\nonumber\\
&=(U^{\dagger}4MM^{
\dagger}U)_{00}-|(U^{\dagger}2MU)_{00}|^2=(\Delta M_0)^2,
\end{eqnarray}
\begin{figure}[b]
  \centering
  \subfigure[][]{
 \includegraphics[width=0.2\textwidth]
{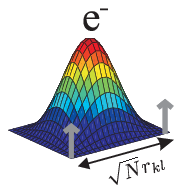}}\label{fig:dipoledipole1}\hfill \subfigure[][]{
 \includegraphics[width=0.35\textwidth]
{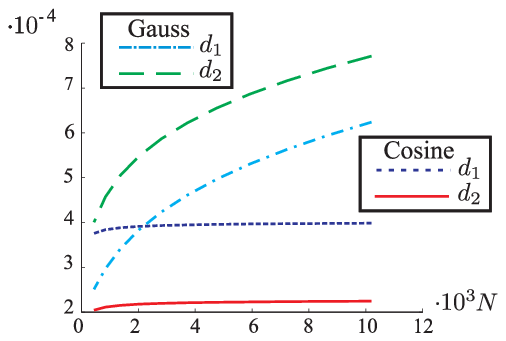}}\hfill  \label{fig:dipoledipole2}
  \subfigure[][]{\includegraphics[width=0.35\textwidth]{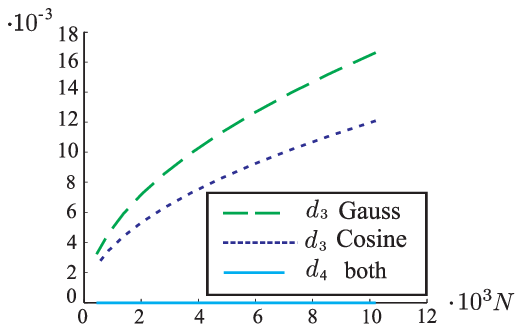}
  }
  \label{fig:dipoledipole3}
  \caption{a) Cosine-shaped wavefunction of the electron on a
  $2$-dimensional
square grid with the nuclear spins located at the vertex points, b)
plot of the ratios $d_1=\frac{\tilde{\gamma}(\Delta M_0)}{g_2}$ and
$d_2=\frac{\tilde{\gamma}(\Delta S_0)}{g_2}$ for a cosine and a
Gaussian shaped wave function. For $N=10^4$, both ratios $d_1$ and
$d_2$ are on the order of $10^{-4}$, together with c), we see that
the dipolar interaction is negligible. c) plot of the ratios
$d_3=\frac{\tilde{\gamma}(U^{\dagger}SU)_{00}}{g_2}$ and
$d_4=\frac{\tilde{\gamma}(U^{\dagger}2MU)_{00}}{g_2}$ for a cosine
and a Gaussian shaped wave function. For $N=10^4$, $d_3$ is on the
order of $10^{-2}$ and $d_4$ is zero due to the symmetry of the
electron wavefunction.}
 \label{fig:dipoledipole}
\end{figure}
and $|c_2|^2=(\Delta S_0)^2$ for the second term. $\Delta M_0$ and
$\Delta S_0$  depend only on the electron wave function and the
lattice geometry. To numerically calculate $\Delta M_0$ and $\Delta
S_0$ and the effect of the last two terms in
(\ref{eq:dipoledipole}), we consider the case where the nuclei lie
in a $2$-dimensional square plane with length $R=\sqrt{N}r_{0}$ of
each side
 on a grid with equal spacings $r_{0}$ (=$0.24nm$
in GaAs \cite{SKL03})[see figure \ref{fig:dipoledipole}(a)].
Consequently, $\theta_{ij}=\pi/2$, which simplifies many expressions
in $H_{dd}$. These assumptions can be made as the height of the QD
is small compared to its diameter, so that the variation of $\theta$
that is dependent of the height of the QD is small,
$\theta_{ij}\approx\pi/2$.

To illustrate our results we consider two simple choices for the
electron wavefunction such that $\alpha_{l}^{(0)}=\frac{1}{\sum_l
f_{1/2}(\mathbf{r}_l)}f_{1/2}(\mathbf{r}_l)$ with
$\mathbf{r}_l=(x_l,y_l)$,
\begin{equation}
f_1(\mathbf{r}_l)=\cos{\left(\frac{\pi}{2}\frac{x_l}{R}\right)}^2\cdot\cos{\left(\frac{\pi}{2}\frac{y_l}{R}\right)}^2,
\end{equation} and
\begin{equation}
f_2(\mathbf{r}_l)=\exp{(-\sqrt{2}r_l^2/R^2)}.
\end{equation}

To show that the direct dipolar interaction is a weak effect
compared to the optical-nuclear coupling $g$, we calculate the
ratios
\begin{eqnarray}
&d_1=\frac{\tilde{\gamma}(\Delta
M_0)}{g_{2}}=\frac{\tilde{\gamma}(\Delta
M_0)}{\frac{\Omega_c\Omega_lg_n}{8\Delta'_{T_{-}}\tilde{\omega}_e}}=\frac{8\Delta'_{T_{-}}\tilde{\omega}_e}{\Omega_c\Omega_l}\frac{\tilde{\gamma}}{A}\frac{(\Delta
M_0)}{\sqrt{\sum_{i=1}^N\alpha_i^{(0)2}}}.
\end{eqnarray}
and $d_2=\frac{\tilde{\gamma}\Delta S_0}{g_2}$. For the parameters
used for the simulation in Section \ref{sec:coupling},
$\frac{8\Delta'_{T_{-}}\tilde{\omega}_e}{\Omega_c\Omega_l}\frac{\tilde{\gamma}}{A}\approx4\cdot10^{-5}$
with $\tilde{\gamma}$ for GaAs \cite{SKL03}. A plot of $d_1$ and
$d_2$ is shown in figure~\ref{fig:dipoledipole}(b). $d_1$ and $d_2$
are both on the order of $10^{-4}-10^{-5}$, for $N>1000$ nuclear
spins and increase slowly with $N$. The last two terms in
(\ref{eq:dipoledipole}), $(U^{\dagger}SU)_{00}b^{\dagger}_0b_{0}$
and $(U^{\dagger}2MU)_{00}b_0b_{0}$ are small and zero,
respectively, as can be seen in figure \ref{fig:dipoledipole}(c):
The ratio of $d_3=\frac{\tilde{\gamma}(U^{\dagger}SU)_{00}}{g_2}$ is
on the order of $10^{-3}-10^{-2}$ for $N>1000$ nuclear spins and the
ratio $d_4=\frac{\tilde{\gamma}(U^{\dagger}2MU)_{00}}{g_2}$ is zero
due to the symmetry of the electron wavefunction in this setting.
Shifting the electron wavefunction such that it is not longer
symmetric with respect to the coordinate origin, $d_4$ is on the
order of $10^{-4}$. We assume that the nuclei lie in a plane, so
there is no displacement of $b_0$ as $D_{ij}=0$ for
$\theta_{ij}=\frac{\pi}{2}$. Therefore, we have shown, that direct
dipolar coupling is an effect that does not affect our protocol.

The hyperfine coupling between electron spin and nuclear spins leads
to a \textit{mediated} dipolar interaction between nuclear spins
\cite{YLS06}. In the bosonic description, the electron couples
solely to the $b_0$ mode, thus, the mediated coupling leads only to
an energy shift
\begin{equation}
\frac{g_n^2}{4\tilde{\omega}_e}b^{\dagger}_0b_0
\end{equation}
that depends on the Zeeman splitting $\tilde{\omega}_e$ and the
number of nuclear excitations. This was already present in
(\ref{eq:Heff1b}) and is not affecting the protocol, in fact it can
help as Kurucz et al. \cite{KuFl08} showed.

For spin-$1/2$ systems, as considered here, the quadrupolar
interaction is not present. For large spin I (e.g. $3/2$  or $9/2$)
nuclei present in GaAs, there is a significant quadrupolar term.
Depending on the strain, up to $g_q\lesssim10^{-2}\mu$eV have been
measured \cite{MKI08}. Therefore, for $I>1/2$, dots with small
strain have to be considered. The quadrupolar interaction
\cite{Sli90} can be treated on a similar footing as the dipolar
coupling in Section \ref{sec:dipoledipole}.

\subsection{Errors in the bosonic picture}\label{sec:errorbos}
We have relied on a simple bosonic description of the collective
nuclear excitations and neglected all corrections to that simplified
picture. For homogeneous coupling ($\alpha_j= $const) this is the
well-known Holstein-Primakoff approximation \cite{HoPr40} and for
systems cooled to a dark state \cite{IKTZ03} at moderate
polarization ($\langle A^z\rangle$ on the order of $-1/2$) spin,
replacing the collective spin operators by bosonic operators is
accurate to $o(1/N)$. The generic inhomogeneous case is discussed in
detail in \cite{Christ2008}. In that case, the Hamiltonian
(\ref{eqn:hfbosonic}) can be seen as a zeroth order approximation in
a small parameter $\sim q(1-P)$, where $q\geq1/2$ and $q=1/2$ for a
homogeneous wave function. The first-order correction analyzed in
\cite{Christ2008} contains two contributions: (i) a polarization
dependent scaling of the coupling-strength $g_n$ which has
negligible effect on the adiabatic transfer we consider and (ii) an
effective coupling of $b$ to bath modes due to the inhomogeneity of
the $A^z$ term. This correction can be computed similarly to the one
in the preceding subsections by rewriting $A^z$ in terms of bosonic
operators. The coupling strength of the leading term is found to be
$\sim A/N=g_n/\sqrt{N}$ and is thus much weaker than $g_{1/2}$.
Since $g_{1/2}$ also characterizes the energy splitting between
different excitation-manifolds in the JC system, this term is
further suppressed by energy considerations.

\section{Summary and Conclusions}
We have shown how to realize a quantum interface between the
polarized nuclear spin ensemble in a singly charged quantum dot and
a traveling optical field. The coupling is mediated by the electron
spin and the mode of a high-Q optical cavity to which the quantum
dot is strongly coupled. Our proposal exploits the strong hyperfine
and cavity coupling of the electron to eliminate the electronic
degree of freedom and obtain an effective coupling between cavity
and nuclei. First, we have studied several possibilities to directly
map the state of the cavity to the nuclei and discussed error
processes and drawbacks of these schemes. Then, we have presented a
more sophisticated interface which is robust to cavity decay.
Read-out is achieved via cavity decay while write-in is based on the
generation of two-mode squeezed states of nuclei and output field
and teleportation. For typical values of hyperfine interaction and
cavity lifetimes, several ebit of entanglement can be generated
before internal nuclear dynamics becomes non-negligible. All
proposed schemes take advantage of the bosonic character of the
nuclear system at high polarization, which implies that all the
relevant dynamics of nuclei, cavity and output field is described by
quadratic interactions. This allows the analytical solution of the
dynamics and a detailed analysis of the entanglement generated. We
show that apart from mapping a light state to the nuclei, the
couplings described enable the preparation of arbitrary Gaussian
states of the nuclear mode.

For highly polarized nuclear spin systems the bosonic description
provides a very convenient framework for the discussion of (dipolar
and quadrupolar) ``internal'' nuclear dynamics. It is seen that
these processes do not appreciably affect the performance of the
interface.

Our results give further evidence that nuclear spins in quantum dots
can be a useful system for quantum information processing. In view
of the recent impressive experimental progress in both dynamical
nuclear polarization of quantum dots and quantum dot cavity-QED,
their use for QIP protocols may not be too far off.

\section{Acknowledgements} This work was supported by the DFG within
SFB 631 and the Excellence Cluster NIM.

\appendix

\section{Gaussian states and operations}\label{Appendix1}

Gaussian states and operations play a central role in quantum
information with continuous variable systems \cite{BrLo05}. To make
this work self-contained we briefly summarize here the main
properties of Gaussian states and operations with particular regard
to their entanglement.

Gaussian states are a family of states occurring very frequently in
quantum optics, e.g., in the form of coherent, squeezed, and thermal
states. Despite being defined on an infinite dimensional Hilbert
space [${\cal
  F}_+(\mathbbm{R}^{2N})$, the symmetric Fock space over $\mathbbm{R}^{2N}$] they
are characterized by a finite number of real parameters, namely the
first and second moments of $N$ pairs of canonically conjugate
observables $(Q_1,P_1,\dots,Q_N,P_N)\equiv\vec{R}$.

One way to define them is that their \emph{characteristic function},
i.e., the expectation values $\chi(\xi)=\mathrm{tr}(W_\xi\rho)$ of
the displacement operators $W_\xi=\exp(i\xi^T\vec{R}),
\xi\in\mathbbm{R}^{2N}$ is a Gaussian function \cite{MaVe68}:
\begin{equation}\label{eq:chi}
\chi(\rho)=\exp(-i\xi^T d-1/4\xi^T\gamma\xi).
\end{equation}
The displacement vector $d\in\mathbbm{R}^{2N}$ and the $2N\times 2N$
real
 positive covariance matrix (CM) $\gamma$ are given by the expectations and
 (co)variances of the $R_k$:
\numparts
\begin{eqnarray}
   \label{eq:2}
     d_k&=&\mr{tr}[\rho R_k],\\
\gamma_{kl}&=&\langle R_i R_j+R_jR_i\rangle-2\langle
R_i\rangle\langle R_j\rangle.
\end{eqnarray}\endnumparts
All $d\in\RR^{2N}$ are admissible displacement vectors and any real
positive matrix $\gamma$ is a valid CM if it satisfies $\gamma\geq
i\sigma_N$ when the symplectic matrix $\sigma_N$ is
\begin{equation}
  \label{eq:sigma}
  \sigma_N =
\oplus_{l=1}^N\sigma_1\,\,\, \textrm{with}\,\,\, \sigma_1 = \left( \begin{array}{cc} 0&-1\\
    1&0 \end{array} \right).
\end{equation}
The last condition summarizes all the uncertainty relations for the
canonical operators $R_j$. These operators are related to the
creation and annihilation operators $a^{\dagger}_j,a_j$ by the
relations $Q_j=(a_j+a_j^{\dagger})/\sqrt{2}$ and
$P_j=-i(a_j-a_j^{\dagger})/\sqrt{2}$.

An example for a one-mode Gaussian state is a coherent state
$\ket{\alpha}$, with covariance matrix $\gamma=\openone$ and
displacement $d=(\Re|\alpha|,\Im|\alpha|)/\sqrt{2}$.

\emph{Entanglement: } All information about the entanglement
properties of Gaussian states is encoded in the CM. Given a CM,
there are efficient criteria to decide whether a Gaussian state is
entangled or not.

To apply these criteria, it is useful to write the CM of a bipartite
$N\times M$ Gaussian states in the following form,
\begin{eqnarray}\label{eq:cov}
\gamma= \left(\begin{array}{cc}
A&C\\
C^T&B
\end{array}\right),
\end{eqnarray}
where the $2N\times 2N$ ($2M\times 2M$) matrix $A$ ($B$) refers to
the covariances of the quadrature operators associated with the
first (second) system and $C$ contains the covariances between the
two systems. $A$ ($B$) are the CM of the reduced state in the first
(second) system only.

In the case of a two-mode system the criteria \cite{DGCZ00,Sim00}
are necessary and sufficient for separability: a state with CM
$\gamma$ is entangled if and only if $\det\gamma + 1-\det A-\det
B+2\det C\not\geq 0$. In this case, entanglement is necessarily
accompanied by a non-positive partial transpose (npt) \cite{Per96}.
For more modes, entangled states with positive partial transpose
exist \cite{WeWo01} and more general criteria to decide entanglement
have to be used \cite{GKLC01,HyEi06}.

For pure states, the analysis of entanglement properties becomes
particularly easy since all such states can be transformed to a
simple standard form, namely a collection of two-mode squeezed
states (TMSS) and vacuum states, by local unitaries \cite{GECP03},
hence the entanglement of such a state is fully characterized by the
vector of two-mode squeezing parameters.  This also shows that for a
$1\times M$ system in a pure state one can always identify a single
mode such that only it (and not the $M-1$ other modes) is entangled
with the first system.

For many Gaussian states it is also possible to make
\emph{quantitative} statements about the entanglement, i.e. to
compute certain entanglement measures. For pure $N\times M$ states,
the entropy of entanglement can be computed from the symplectic
eigenvalues of the reduced CM $A$ (or, equivalently, $B$). These are
given by the modulus of the eigenvalues of $\sigma_N A$
\cite{ViWe01}. All symplectic eigenvalue $\lambda\geq1$ corresponds
to a TMSS with squeezing parameter $\acosh(\lambda)/2$ in the
standard form of the state at hand and contributes
$\lambda^2\log_2\lambda^2-(\lambda-1)^2\log_2(\lambda^2-1)$ to the
entanglement entropy of the system. \\
For mixed states, it is possible to compute the negativity
\cite{ViWe01} for any $N\times M$ system from the symplectic
eigenvalues of the CM of the partially transposed state (which is
related to the CM obtained by replacing all momenta $P_j$ in the
second system by $-P_j$). Every symplectic eigenvalue
$\lambda<1$ contributes $-\log_2\lambda$ to the negativity. \\
For $1\times1$ Gaussian states with $\det A=\det B$ (so-called
symmetric states), the entanglement of formation (EoF) can be
computed \cite{GWK+03} and for more general states a Gaussian
version of EoF is available \cite{WGK+03}. Even if the states are
not certain to be Gaussian, several of the Gaussian quantities can
serve as lower bounds for the actual amount of entanglement
\cite{WGC06}.

\emph{Gaussian operations: } Operations that preserve the Gaussian
character of the states they act on are called Gaussian operations
\cite{GiCi02}. Like the Gaussian states they are only a small family
(in the set of all operations) but play a prominent role in quantum
optics, since they comprise many of the most readily implemented
state transformations and dynamics. With Gaussian operations and
Gaussian states many of the standard protocols of quantum
information processing such as entanglement generation, quantum
cryptography, quantum error correction and quantum teleportation can
be realized \cite{BrLo05}.

Of particular interest for us are the Gaussian unitaries, i.e.
unitary evolutions generated by Hamiltonians that are at most
quadratic in the creation and annihilation operators. Unitary
displacements $W_\xi$ are generated by the linear Hamiltonian
$\xi^T\vec{R}$. All other Gaussian unitaries can be composed of
three kinds \cite{Bra99}, named according to their optical
incarnations. The \emph{phase shifter} ($H=a^{\dagger}a$)
corresponds to the free evolution of an harmonic system. The
\emph{beam
  splitter} ($H=ab^\dagger + a^\dagger b$) couples two modes. Both generators
do not change the total photon number and are therefore examples of
\emph{passive} transformations. The remaining type of Gaussian
unitary is \emph{active}: the (single-mode) \emph{squeezer} is
generated by the squeezing Hamiltonian $H=a^2+(a^\dagger)^2$, which,
when acting on the vacuum state decreases the variance in one
quadrature ($Q$) by a factor $f<1$ and increases the other one by
$1/f$. Combining these building blocks in the proper way, all other
unitaries generated by quadratic Hamiltonians, e.g. the two-mode
squeezing transformation ($H=ab+a^\dagger b^\dagger$) can be
obtained.

Both active and passive transformations map field operators to a
linear combination of field operators (disregarding displacements
caused by linear parts in the Hamiltonians, which can always be
undone by a further displacement), i.e. for all Gaussian unitaries
we have in the Heisenberg picture
\begin{equation}
U\vec{R} U^\dagger = S\vec{R} \equiv \vec{R}'.
\end{equation}
Here $S$ is a symplectic map on $\mathbbm{R}^{2N}$, i.e. $S$
preserves the symplectic matrix $\sigma_N$, assuring that $R_i$ and
$R_i'$ satisfy the same commutation relations. We denote by $U_S$
the unitary corresponding to the symplectic transformation $S$.
Passive operations correspond to symplectic transformations that are
also orthogonal.

In the Schr\"odinger picture, $U_S$ transforms the Gaussian state
with CM $\gamma$ and displacement $d$ such that
$(\gamma,d)\mapsto(S\gamma S^T,Sd)$. The two-mode squeezing
transformations
\begin{equation}
  \label{twomodesqu}
  T(r) =
\left(\begin{array}{cc}
  \cosh{(r)}\openone&\sinh{(r)}\sigma_x\cr
\sinh{(r)}\sigma_x& \cosh{(r)}\openone
\end{array}\right)
\end{equation}
used in Sec.~\ref{sec:lzfock} is an important example of a active
symplectic transformation.

Besides Gaussian unitaries, \emph{Gaussian measurements} are another
important and readily available tool. Gaussian measurements are
generalized measurements represented by a positive-operator-valued
measure $\left\{
  \proj{\gamma,d}{\gamma,d}, d\in\mathbbm{R}^{2N} \right\}$ that is formed by all
the projectors obtained from a pure Gaussian state
$\proj{\gamma,0}{\gamma,0}$ by displacements. The most important
example is a limiting case of the above: the quadrature measurements
(von Neumann measurements which project on the (improper, infinitely
squeezed) eigenstates of, e.g., $Q$). In quantum optics, these are
well approximated by \emph{homodyne detection}. For example, the
``Bell- or ``EPR-measurement'' that is part of the teleportation
protocol is a measurement of the commuting quadrature operators
$Q_1+Q_2$ and $P_1-P_2$.

\section{Landau-Zener transitions}\label{Appendix}
In a rotating frame with $U=\exp{[-\frac{i}{2}\int
(\omega_1-\omega_2)\sigma_z+(\omega_1+\omega_2)\openone\,
\textrm{dt}]}$ the Heisenberg equations, given by (\ref{eq:lz}),
read:
\begin{eqnarray}
\dot{u'}=&-i g \exp{\left(i\int (\omega_1-\omega_2)\textrm{dt}\right)}v'\label{eqn:dgl}\\
\dot{v'}=&-i g \exp{\left({-i\int(\omega_1-\omega_2)}
\textrm{dt}\right)}u'\label{eqn:dgl2}
\end{eqnarray}
The initial boundary conditions of the coupled differential
Equations (\ref{eqn:dgl}) and (\ref{eqn:dgl2}) are now chosen such,
that the photon operator $a$ at time $t\rightarrow -\infty$ is
mapped to the nuclear spin operator $b$ at $t\rightarrow \infty$
\begin{eqnarray}\label{eqn:ini1}
u'_{-\infty}=1, |v'_{-\infty}|=0.
\end{eqnarray}
Eliminating $u'$ in (\ref{eqn:dgl}) and (\ref{eqn:dgl2}) leads to
the single equation:
\begin{equation}\label{eqn:landau3}
\ddot{v'}+i\beta t\dot{v'}+g^2v'=0,
\end{equation}
where $\dot{g}=0$. Together with the substitution
$v'=e^{-\frac{i}{2}\int(\omega_1-\omega_2)dt}U_1$,
(\ref{eqn:landau3}) reduces to the so called Weber equation:
\begin{equation}\label{eqn: weber}
\ddot{U_1}+\left(g^2-i\frac{\beta}{2}+\frac{\beta^2}{4}t^2\right)U_1=0.
\end{equation}
Solving (\ref{eqn: weber}) as proposed by Landau and Zener and
considering the asymptotic behavior of the solution at
$t\rightarrow\infty$, it is found to be
\begin{equation}
\lim_{t\to\infty}U_1(t)=-K\frac{\sqrt{2\pi}}{\Gamma(i\gamma_z+1)}e^{-\frac{1}{4}\pi
\gamma_z}e^{i\beta t^2}(\sqrt{\beta}t)^{i\gamma_z},
\end{equation}
where $\gamma_z=\frac{g_1^2}{\beta}$
 and the constant
$K=\sqrt{\gamma_z}\exp{\left(-\frac{\gamma_z\pi}{4}\right)}$. The
probability that the photonic operator $a$ is mapped to the
collective nuclear spin operator $b$ is given by (\ref{eqn:ze}).


\end{document}